\documentclass[12pt, a4paper, makeidx, oneside]
{memoir}

\usepackage{graphicx}
\usepackage{epsfig}
\usepackage{amsmath}
\usepackage{amssymb}
\usepackage{amsthm}
\usepackage{booktabs}
\usepackage{url}
\usepackage{longtable}
\usepackage{lscape}
\usepackage{rotating}
\makeindex

\usepackage{color}

\definecolor{greenyellow}   {cmyk}{0.15, 0   , 0.69, 0   }
\definecolor{yellow}        {cmyk}{0   , 0   , 1   , 0   }
\definecolor{goldenrod}     {cmyk}{0   , 0.10, 0.84, 0   }
\definecolor{dandelion}     {cmyk}{0   , 0.29, 0.84, 0   }
\definecolor{apricot}       {cmyk}{0   , 0.32, 0.52, 0   }
\definecolor{peach}         {cmyk}{0   , 0.50, 0.70, 0   }
\definecolor{melon}         {cmyk}{0   , 0.46, 0.50, 0   }
\definecolor{yelloworange}  {cmyk}{0   , 0.42, 1   , 0   }
\definecolor{orange}        {cmyk}{0   , 0.61, 0.87, 0   }
\definecolor{burntorange}   {cmyk}{0   , 0.51, 1   , 0   }
\definecolor{bittersweet}   {cmyk}{0   , 0.75, 1   , 0.24}
\definecolor{redorange}     {cmyk}{0   , 0.77, 0.87, 0   }
\definecolor{mahogany}      {cmyk}{0   , 0.85, 0.87, 0.35}
\definecolor{maroon}        {cmyk}{0   , 0.87, 0.68, 0.32}
\definecolor{brickred}      {cmyk}{0   , 0.89, 0.94, 0.28}
\definecolor{red}           {cmyk}{0   , 1   , 1   , 0   }
\definecolor{orangered}     {cmyk}{0   , 1   , 0.50, 0   }
\definecolor{rubinered}     {cmyk}{0   , 1   , 0.13, 0   }
\definecolor{wildstrawberry}{cmyk}{0   , 0.96, 0.39, 0   }
\definecolor{salmon}        {cmyk}{0   , 0.53, 0.38, 0   }
\definecolor{carnationpink} {cmyk}{0   , 0.63, 0   , 0   }
\definecolor{magenta}       {cmyk}{0   , 1   , 0   , 0   }
\definecolor{violetred}     {cmyk}{0   , 0.81, 0   , 0   }
\definecolor{rhodamine}     {cmyk}{0   , 0.82, 0   , 0   }
\definecolor{mulberry}      {cmyk}{0.34, 0.90, 0   , 0.02}
\definecolor{redviolet}     {cmyk}{0.07, 0.90, 0   , 0.34}
\definecolor{fuchsia}       {cmyk}{0.47, 0.91, 0   , 0.08}
\definecolor{lavender}      {cmyk}{0   , 0.48, 0   , 0   }
\definecolor{thistle}       {cmyk}{0.12, 0.59, 0   , 0   }
\definecolor{orchid}        {cmyk}{0.32, 0.64, 0   , 0   }
\definecolor{darkorchid}    {cmyk}{0.40, 0.80, 0.20, 0   }
\definecolor{purple}        {cmyk}{0.45, 0.86, 0   , 0   }
\definecolor{plum}          {cmyk}{0.50, 1   , 0   , 0   }
\definecolor{violet}        {cmyk}{0.79, 0.88, 0   , 0   }
\definecolor{royalpurple}   {cmyk}{0.75, 0.90, 0   , 0   }
\definecolor{blueviolet}    {cmyk}{0.86, 0.91, 0   , 0.04}
\definecolor{periwinkle}    {cmyk}{0.57, 0.55, 0   , 0   }
\definecolor{cadetblue}     {cmyk}{0.62, 0.57, 0.23, 0   }
\definecolor{cornflowerblue}{cmyk}{0.65, 0.13, 0   , 0   }
\definecolor{midnightblue}  {cmyk}{0.98, 0.13, 0   , 0.43}
\definecolor{navyblue}      {cmyk}{0.94, 0.54, 0   , 0   }
\definecolor{royalblue}     {cmyk}{1   , 0.50, 0   , 0   }
\definecolor{blue}          {cmyk}{1   , 1   , 0   , 0   }
\definecolor{cerulean}      {cmyk}{0.94, 0.11, 0   , 0   }
\definecolor{cyan}          {cmyk}{1   , 0   , 0   , 0   }
\definecolor{processblue}   {cmyk}{0.96, 0   , 0   , 0   }
\definecolor{skyblue}       {cmyk}{0.62, 0   , 0.12, 0   }
\definecolor{turquoise}     {cmyk}{0.85, 0   , 0.20, 0   }
\definecolor{tealblue}      {cmyk}{0.86, 0   , 0.34, 0.02}
\definecolor{aquamarine}    {cmyk}{0.82, 0   , 0.30, 0   }
\definecolor{bluegreen}     {cmyk}{0.85, 0   , 0.33, 0   }
\definecolor{emerald}       {cmyk}{1   , 0   , 0.50, 0   }
\definecolor{junglegreen}   {cmyk}{0.99, 0   , 0.52, 0   }
\definecolor{seagreen}      {cmyk}{0.69, 0   , 0.50, 0   }
\definecolor{green}         {cmyk}{1   , 0   , 1   , 0   }
\definecolor{forestgreen}   {cmyk}{0.91, 0   , 0.88, 0.12}
\definecolor{pinegreen}     {cmyk}{0.92, 0   , 0.59, 0.25}
\definecolor{limegreen}     {cmyk}{0.50, 0   , 1   , 0   }
\definecolor{yellowgreen}   {cmyk}{0.44, 0   , 0.74, 0   }
\definecolor{springgreen}   {cmyk}{0.26, 0   , 0.76, 0   }
\definecolor{olivegreen}    {cmyk}{0.64, 0   , 0.95, 0.40}
\definecolor{rawsienna}     {cmyk}{0   , 0.72, 1   , 0.45}
\definecolor{sepia}         {cmyk}{0   , 0.83, 1   , 0.70}
\definecolor{brown}         {cmyk}{0   , 0.81, 1   , 0.60}
\definecolor{tan}           {cmyk}{0.14, 0.42, 0.56, 0   }
\definecolor{gray}          {cmyk}{0   , 0   , 0   , 0.50}
\definecolor{black}         {cmyk}{0   , 0   , 0   , 1   }
\definecolor{white}         {cmyk}{0   , 0   , 0   , 0   } 

\ifpdf
        \usepackage[plainpages=false,pdfpagelabels,bookmarksnumbered,%
        colorlinks=true,%
        linkcolor=sepia,%
        citecolor=sepia,%
        filecolor=maroon,%
        pagecolor=red,%
        urlcolor=sepia,%
        pdftex,%
        unicode]{hyperref} 
    \usepackage{thumbpdf} 
\else
    \usepackage{hyperref}
\fi

\usepackage{memhfixc}

\settypeblocksize{*}{33pc}{1.618}

\setulmarginsandblock{4cm}{3cm}{*}
\setlrmarginsandblock{2.8cm}{2.5cm}{*}

\setheadfoot{\onelineskip}{2\onelineskip}
\setheaderspaces{*}{2\onelineskip}{*}

\checkandfixthelayout

\makechapterstyle{mychapterstyle}{%
}

\chapterstyle{mychapterstyle}

\setsecheadstyle{\Large\sffamily\bfseries}
\setsubsecheadstyle{\large\sffamily\bfseries}
\setsubsubsecheadstyle{\normalfont\sffamily\bfseries}
\setparaheadstyle{\normalfont\sffamily}

\makeevenhead{headings}{\thepage}{}{\small\slshape\leftmark}
\makeoddhead{headings}{\small\slshape\rightmark}{}{\thepage}

\settocdepth{subsection}

\setsecnumdepth{subsection}
\maxsecnumdepth{subsection}
\settocdepth{subsection}
\maxtocdepth{subsection}

\begin{document}
\frontmatter


\pagestyle{empty}
\sffamily

\noindent
\begin{center}
    \Large
    Universidade Federal do Rio de Janeiro\\
    Centro de Ci\^encias Matem\'aticas e da Natureza\\
    Observat\'orio do Valongo
\end{center}

\vfill\vfill
\begin{center}
    \large
    Master of Science Dissertation\\
\end{center}

\vfill
\begin{center}
    \Huge\bfseries
    Single Past Null Geodesic in the Lema\^itre-Tolman-Bondi Cosmology
\end{center}

\vfill
\begin{center}
    \Large
    by
\end{center}

\vfill
\begin{center}
    \LARGE\bfseries
    Felipe Antonio Monteiro Gomes Nogueira
\end{center}

\vfill\vfill\vfill
\begin{center}
    \Large
    Supervisor: Marcelo Byrro Ribeiro
\end{center}

\vfill
\begin{center}
\large
    Rio de Janeiro, November 2013
\end{center}
\pagenumbering{arabic}
\setcounter{page}{1}
\clearpage





\pagestyle{empty}
\clearpage


\begin{center}
{\LARGE Felipe Antonio Monteiro Gomes Nogueira}
\par
\vspace{200pt}
{\Huge Single Past Null Geodesic in the Lema\^itre-Tolman-Bondi Cosmology}
\end{center}
\par
\vspace{70pt}
\hspace*{175pt}\parbox{7cm}{{\large Disserta\c c\~ao de mestrado apresentada ao Programa
de P\'os-gradua\c c\~ao em Astronomia do Observat\'orio
do Valongo, Universidade Federal do Rio de
Janeiro, como requisito parcial \`a obten\c c\~ao do
t\'itulo de mestre em Astronomia.}}
\par
\vspace{1em}
\vfill
\begin{center}
{\large Orientador: Marcelo Byrro Ribeiro}\\
\vspace{1em}
\textbf{{\large Rio de Janeiro}\\
{\large Novembro de 2013}}
\end{center}

\newpage

\fbox{\begin{minipage}{13cm}
Nogueira, Felipe A. M. G. \\
\hspace{2em} Single Past Null Geodesic in the Lema\^itre-Tolman-Bondi Cosmology \\
\hspace{2em} f: 72f. \\
\hspace{2em} Orientador: Marcelo Byrro Ribeiro \\
\hspace{2em} Disserta\c c\~ao (Mestrado em Astronomia) - UFRJ/OV - Programa de P\'os-gradua\c c\~ao em Astronomia, 2013. \\
Refer\^encias Bibliogr\'aficas: f: 69-72. \\
1.Cosmologia. 2. Relatividade. 3. Fractais.
I. Ribeiro, Marcelo B. II. Universidade Federal do Rio de Janeiro. Observat\'orio do Valongo. Programa de P\'os-Gradua\c c\~ao em Astronomia. III. T\'itulo.
\end{minipage}}
\par
\vspace{2em}







\newpage

\vspace*{0.5\textheight}
\noindent \textit{``Entia non sunt multiplicanda praeter necessitatem."}\footnote[1]{Entities must not be multiplied beyond necessity.}
\\
William of Ockham, \textit{Lex Parsimoniae}
\\
\\
\textit{``An important scientific innovation rarely makes its way rapidly winning over and converting its opponents; it rarely happens that Saul becomes Paul. What does happen is that its opponents gradually die out and that the growing generation is familiarized with the idea from the beginning."}
\\
Max Planck, The Philosophy of Physics (1936)
\\
\\
\textit{``What we observe is not nature itself, but nature exposed to our method of questioning."}
Werner Heisenberg, Physics and Philosophy: The Revolution in Modern Science (1958)

\clearpage


\noindent{\LARGE\textbf{Acknowledgments}}
\\
\\

First of all I want to thank my family, who supported me throughout these years, without whom this would not be possible. \\

I also want to thank all my friends who shared the good and the not so good moments with me, including my colleagues at OV, my friends from my two homes, Niter\'oi and Araruama, and many others who participated in this quest somehow. \\

A big thank you to my very patient supervisor Dr. Marcelo Byrro Ribeiro, who has been a great teacher and helped me in almost every aspect of becoming a reseacher. \\

Finally I would like to thank CAPES for the financial support over these two years.

\chapter*{Abstract}

This work provides a general discussion of the spatially inhomogeneous Lema\^itre-Tolman-Bondi (LTB) cosmology, as well as its basic properties and many useful relevant quantities, such as the cosmological distances. We apply the concept of the single null geodesic to produce some simple analytical solutions for observational quantities such as the redshift. As an application of the single null geodesic technique, we carry out a fractal approach to the parabolic LTB model, comparing it to the spatially homogeneous Einstein-de Sitter cosmology. The results obtained indicate that the standard model, in this case represented by the Einstein-de Sitter cosmology, can be equivalently described by a fractal distribution of matter, as we found that different single fractal dimensions describe different scale ranges of the parabolic LTB matter distribution. It is shown that at large ranges the parabolic LTB model with fractal dimension equal to 0.5 approximates the matter distribution of the Einstein-de Sitter universe.


\chapter*{Resumo}

Este trabalho fornece uma discuss\~ao geral da cosmologia espacialmente inomog\^enea de Lema\^itre-Tolman-Bondi (LTB), bem como de suas propriedades b\'asicas e algumas grandezas \'uteis, como as dist\^ancias cosmol\'ogicas. A partir da aplica\c c\~ao do conceito de geod\'esica nula \'unica, obtivemos solu\c c\~oes anal\'iticas simples para grandezas relevantes como o \textit{redshift}. Como aplica\c c\~ao deste conceito de geod\'esica nula \'unica, desenvolvemos uma abordagem fractal para o modelo inomog\^eneo de LTB, comparando-o com a cosmologia homog\^enea de Einstein-de Sitter. Os resultados indicam que o modelo padr\~ao, neste caso representado pela cosmologia de Einstein-de Sitter, pode ser descrito de forma equivalente a uma distribui\c c\~ao fractal de mat\'eria, pois encontramos que dimens\~oes fractais diferentes descrevem regi\~oes diferentes do modelo parab\'olico de LTB. \'E mostrado que a grandes escalas o modelo de LTB parab\'olico com dimens\~ao fractal igual a 0.5 aproxima-se da distribui\c c\~ao de mat\'eria do universo do modelo de Einstein-de Sitter.


\clearpage
\pagestyle{headings}
\tableofcontents*
\clearpage
\listoffigures

\mainmatter
\setcounter{page}{12}
\chapter*[Introduction]{Introduction}
\addcontentsline{toc}{chapter}{Introduction}
\label{Introduction}

\indent The Lema\^itre-Tolman-Bondi (LTB) geometry is a spatially inhomogeneous description of a spherically symmetric distribution of matter in the Universe. It is generally seen as an alternative to the usual Friedmann-Lema\^itre-Robertson-Walker (FLRW) spatially homogeneous and isotropic geometry. The LTB model has three arbitrary functions - that can be reduced to two functions by a coordinate transformation - which can be defined in a way that they may obey some theoretical and/or observational requirements. The model has been extensively studied since it was first obtained by A. Georges Lema\^itre (1933), re-derived and interpreted by Richard C. Tolman (1934)  and H. Bondi (1947), and rediscovered by many other authors since then (see Krasi\'nski 1997 and references therein). As we shall see below, there are many different interpretations of the LTB geometry which, although different, usually do not conflict with each other, but give complementary meanings.

The first of these interpretations was given by Lema\^itre when he first found the inhomogeneous solution in 1933. He investigated the formation and condensation of the galaxies by using a small deviation from the Einstein static solution of the field equations. His condensation model proved to be wrong, as it could not predict the galaxy formation by utilizing a point-source model with a non-zero cosmological constant, because the generated perturbations were too small (Bonnor 1956), but the solution remained as a new model.

In the following year, Tolman (1934) obtained a solution for the field equations that described the matter in the Universe as inhomogeneously distributed. He claimed that the reason for obtaning those solutions came from the fact that extrapolations were being made on the basis of the homogeneous models considered at the time, such as the use of heuristic arguments of mathematical simplicity for the use of homogeneous solutions and that the study of inhomogeneities over long periods of time and in very distant regions would have to be made so that premature conclusions could be avoided. His dust, zero pressure spherically symmetric model is the basis of all contemporary LTB models, such as the recently proposed LTB void models (Garcia-Bellido \& Haugb\o lle 2008).

In 1947 H. Bondi obtained the same solution as Tolman's, and studied the Doppler shift as a very interesting feature of the model. He interpreted the equation of motion - found by Tolman as well -, as an energy equation, like the Newtonian energy equation for the gravitational potential, and even showed that the model could be used to obtain other specific cosmological solutions by making suitable choices of its arbitrary functions.

Since then the LTB model has been studied by many authors. A work of particular great significance is the one advanced by W. B. Bonnor (1972, 1974). In his papers he presented a different and simpler notation for the LTB metric, showed that fixing one of the functions of the model could considerably simplify the expressions, generate different solutions, and demonstrated that the LTB cosmology could evolve to a homogeneous Einstein-de Sitter universe when certain conditions are fulfilled.

All these works and applications formed an initially solid theoretical ground for the inhomogeneous LTB cosmology, being a more general metric than the common FLRW models that are widely used nowadays\footnote[1]{Alfadeel \& Hellaby (2010) studied an even more general metric, the Lema\^itre space-time, where pressure is non-zero. In this metric it is even more difficult to define the mass in an invariant way, but, on the other hand, it could offer more flexibility and the possibility of fitting even more data than the LTB geometry.}. Nevertheless, due to its generality, the LTB spacetime comes with a recurring problem of finding analytical solutions to the relevant observational quantities because the null geodesic equations are very difficult to solve analytically. Due to that, Ribeiro (1992ab, 1993) performed a numerical approach to find approximate solutions to the null geodesic equations. In another attempt, Stoeger et al. (1992) showed that changing the traditional LTB coordinates to what they called as `observational coordinates' (Ellis et al. 1985) led to solvable null geodesic equations. But, as it turned out, in this treatment the Eistein field equations became unsolvable analytically. In an attempt to avoid this problem, Mustapha et al. (1997) used a chosen single null geodesic in order to find an analytical solution for the studied quantities. They reasoned that there is essentially one event when astronomical observations of cosmological relevance are made, then there is no need of a general solution for the null geodesic equation. A draw-back is the impossibility of studying the model's evolution.

The LTB spacetime itself assumes inhomogeneity and isotropy relative to a certain point in the Universe. This aspect of the LTB geometry contradicts the common idea that is usually taken as truth in the cosmology community for the those who work with the standard model of cosmology: the homogeneity and isotropy of the Universe, or the Cosmological Principle, which states that the Universe has the same aspect if observed at any point.  

We can say that there is a philosophical reason for this hypothesis being so hard to drop. It is a very simple way of describing the Universe, and simplicity is almost like a principle in physical theories. This simplicity in the description of the Universe leads to the FLRW models commonly used, which have been sucessful in explaining many observed aspects of the Universe, the most notable being the cosmic microwave background (CMB) radiation. Nonetheless, for the FLRW models to generate the observed inhomogeneities and the consequent formation of galaxies, clusters, superclusters and the filaments of matter observed in the large scale structure of the Universe and explain the possible accelerated expansion, other complementary initial condition or mechanisms are needed. That being said, the argument of simplicity cannot hold if several complementary mechanisms are needed to make the theory solid and functionable. Among other possibilities, such as the anisotropic Bianchi models, which may also consider cosmological dissipative fluids to account for the anisotropies of the CMB radiation (Misner 1968, Ribeiro 1987, Pleba\'nski \& Krasi\'nski 2006), a change in the metric that describes the spacetime may be simpler than adding new features to a cosmological model, additions which can be thought of as equivalent to modern day Ptolomaic epicycles.

Among the observational features which may require a change in the spacetime metric so that such features are properly characterized, it is worth mentioning that observations of the large scale structure of the Universe often show that the distribution of galaxies is concentrated in groups, which one can think, among other possibilities, to be hierarchically organized. The hierarchical concept means that the matter in the Universe is grouped in self-similar structures, namely galaxies, clusters and super-clusters of galaxies, depending on the scales considered.

The concept of self-similar structures is well known, and can be called as a fractal system (Mandelbrot 1982). The idea of fractals for describing the observed structures in the Universe was transformed into a mathematical concept by Mandelbrot (1982) and was studied by Pietronero (1987) in a Newtonian viewpoint, where he presented a very simple model in which he defined the fractal distribution by a power law relating the number of objects to a certain distance. Such a relation is bound to divide the universe in two different regions, the one that belongs to the fractal system defined by that relation and the other one that does not. This scenario relates to a discrete geometry, where there are different locally homogeneous regions of the Universe (local isotropy), a weaker version of the Cosmological Principle, stated by Mandelbrot (1982) and called as the Conditional Cosmological Principle. The matter density would be then coordinate (spatially) dependent, meaning that the Universe would be, in this model, described by an inhomogeneous distribution of matter that would be compatible with a fractal distribution of matter.

This simple particular scenario, shared by some authors in the 1990s (Coleman \& Pietronero 1992, Sylos Labini 1998), has changed over time. The problem can be approached from a different viewpoint: observations are made along the past null cone of the observer. Then, observed inhomogeneities are natural consequences of a slicing process of the spacetime, when a specific instant (redshift) of the Universe is considered. Then the inhomogeneity would be an observational feature of the spatially homogeneous universe in this interpretation.

As will be further explained, the arbitrariness of the LTB geometry and its inhomogeneous characteristic can be used for modelling a fractal cosmology. Such an approach amounts to a hierarchical fractal model which has been studied by other authors (Ribeiro 1992ab, Ribeiro 1993, Sylos-Labini 1998, 2011). In the present work, the LTB spacetime is used to advance a fractal approach for describing the distribution of matter in the Universe, following the work of Ribeiro (1992ab, 1993), but differing from it in the sense that this author performed a numerical approach to solve the null geodesic equations derived from the model. The use of the single null geodesic technique provides analytical relations for the matter distribution, as obtained by observational quantities. As we shall see further, the simplicity of the solutions helps in the understanding and interpretation of the inhomogenous fractal model.

These previous developments were of considerable significance in the LTB fractal model that is presented in this work. As Bonnor (1972) showed, there could be regions of the spacetime where the de Vaucoulers (1970) density relation would be valid, within the inhomogeneous model in a hypersurface of constant $ t $. The de Vancoulers density relation he used, $ \rho \propto  (\mathrm{distance})^{1.7} $, is very similar to the one adopted in this work.

Another very interesting feature of the LTB models is that the modeled accelerated expansion of the Universe by the standard model of cosmology, with the use of the cosmological constant and a postulated unknown ``dark energy", can be interpreted differently. This question was first asked by C\'el\'erier (1999), who discussed the Type Ia supernovae (SNIa) and their use as "standard candles" for distance measurements and the consequent interpretation of a cosmic acceleration in spatially homogeneous models. She proposed that the spatially inhomogeneous LTB model was a good alternative for FLRW cosmologies with a non-zero cosmological constant for fitting the SNIa data, because the inhomogeneities could take into account the SNIa data without the need to assume a cosmic acceleration, if one is satisfied in ruling out the Cosmological Principle. As stated by Mustapha et al. (1997), what is usually done is trying to determine a theory of source evolution taking the FLRW assumption $a$ $priori$. In other words, ``if the demonstration of homogeneity depends on knowing the source evolution, and validation of source evolution theories
depends on knowing the cosmological model is homogeneous,
then neither is proved." Expanding this point of view, the so-called parametrized LTB void models present simple alternatives for the description of the distribution of the sources, and the apparent consequences of this acceleration, such as the distance measures from the SNIa. The LTB void models show that these distances can be explained without the use of an extra dark energy component (Garcia-Bellido \& Haugb\o lle 2008). 

Another explanation for the dark energy problem has been proposed by D. Wiltshire (2013), in a phenomenological treatment of the averaging problem in General Relativity, where it is proposed that the use of cold dark matter and dark energy to explain the different gravity interactions in different scales is an indicative of a ``renormalization of the notion of gravitational mass in
a hierarchy of nonrigid spacetime structures''. His results may be viewed as a similar, or even complementary interpretation of the fractal (hierarchical) models.

These LTB type models have been recently acting as the main alternative explanation for the acceleration problem, although not without their own difficulties, such as the the fitting of the considered void center (in the context of void LTB models) and the BAO data, although it is within 3$\sigma$ confidence level (Marra \& P\"{a}\"{a}kk\"{o}nen 2010). Another reason for studying LTB models nowadays is a change in the scenario for the inhomogeneous models. These models, mainly the LTB one, always happen to return whenever a new problem reveals a deficiency or inconsistency with the current standard model. When a satisfying explanation for the problem is presented, then the LTB and alternative models like it are again discarded for a period of time, not changing their status of alternative models. So, this work has the intention of filling a gap that has been part of the LTB models for some time, that is, to provide simple analytical expressions and also simple explanations for some common issues of the model in the context of the new rise of the LTB inhomogeneous cosmology which recently have been focusing in solving the so-called ``dark energy" problem. 

The plan of the dissertation is as follows. In chapter \ref{chapterone}, the LTB geometry is presented and its main interpretations are discussed. The relevant observational quantities are obtained and the particular cases that can be derived from the model, and often used to compare with similar ones derived in the FLRW models, are shown in details. Following the analytical development initiated in the first chapter, the concept of the single null geodesics is discussed and a new solution of the redshift is obtained in chapter \ref{chapter3}, as well as the resulting simplified equations that are used in the next chapter for obtaning the analytical expressions in the fractal model. In chapter \ref{chapter4}, the fractal approach for describing galaxy distribution is developed, its relation to the inhomogeneous model is also presented and discussed, and the results of the use of the single null geodesic technique in this fractal model are shown. For completeness and enlightment of the subject, a comparison of the fractal models with the Einstein-de Sitter cosmology is made by the end of the chapter, making it clear that such a comparison is necessary to understand the reason for studying fractals in an inhomogeneous cosmology context. Finally, the conclusions are drawn in the last chapter.



\chapter{The Lema\^itre-Tolman-Bondi Geometry}\label{chapterone}

This chapter begins by obtaining the LTB solution for the field equations of General Relativity and discussing its properties. The relevant observational quantities are defined in section 1.2, and some interesting particular cases are obtained and discussed in section 1.3.

\section{The Field Equation Solutions}\label{solutions}

The Einstein Field Equations of General Relativity are given by,
\begin{equation}
	G_{\mu \nu} = \frac{8 \pi G}{c^4}T_{\mu \nu}, \label{einstein_equations}
\end{equation}
\begin{equation}
G_{\mu \nu} = R_{\mu \nu} - \frac{1}{2}g_{\mu \nu}\mathrm{R}, \label{einstein_eq}
\end{equation}
where the greek indexes run from 0 to 3. Here $ G $ and $ c $ are the gravitational constant and the speed of light, respectively. These equations define the relationship between the space-time geometry and the distribution of matter in the Universe, and come from Einstein's search for a gravitational theory to describe the matter dynamics in space-time, which was finally presented in 1916.

The left-hand side of equation (\ref{einstein_equations}) is the
``geometry side", represented by the Einstein tensor $G_{\mu \nu}$
composed by a combination of the Ricci tensor $R_{\mu \nu}$, the Ricci scalar $ \mathrm{R}$, and the metric tensor $g_{\mu \nu}$ (eq. \ref{einstein_eq}). This
side of the equation defines the chosen space-time metric. The right-hand side of the equation (\ref{einstein_equations}) is the ``matter side", defined by the energy-momentum tensor $ T_{\mu \nu} $ which shows what will be interacting with the space-time defined in the left-hand side of the equation.

The resulting equations will show how the space-time is altered by the mass-radiation content when the metric and energy-momentum tensors are specified. So, under symmetry assumptions and a matter-radiation distribution, the equations can be solved so that the result is a description of the space-time itself.

For the LTB model we treat here, the basic assumptions are:

\begin{enumerate}
\item the geometry is spherically symmetric with a central point;
\item the energy-momentum tensor is given by a perfect fluid (zero rotation).
\end{enumerate}
Taking into consideration the first condition, the most general line element in the 3 + 1 foliation form (Pleba\'nski \& Krasi\'nski 2006) is, 
\begin{equation}
\mathrm{d}S^2 = \mathrm{e}^{C(t,r)}\mathrm{d}t^2 - \mathrm{e}^{A(t,r)}\mathrm{d}r^2 - R^2(t,r)(\mathrm{d}\theta^2 + \sin^2{\theta}\mathrm{d}\phi^2), \label{lineel}
\end{equation}
for the coordinates $ (t, r, \theta, \phi)$, where $ C(t,r) $ and $ A(t,r) $ are functions to be determined, and $ R $ is related to the area of the surface ($t =$ constant, $r =$ constant), called the areal radius. Assuming that the space-time coordinates are comoving and synchronous, the velocity field is given by,
\begin{eqnarray}
u^{\alpha} = \mathrm{e}^{-C/2}{\delta^{\alpha}}_0, \label{velocityfield}
\end{eqnarray}
where $u^{\alpha}$ is the 4-velocity and ${\delta^{\alpha}}_0$ is the delta Kronecker function.  The second assumption above leads to the perfect fluid energy-momentum diagonal tensor, defined as $T_{\mu \nu} = (\rho, -p, -p, -p)$, where $ \rho $ and $p$ are the mass density and pressure, respectively. The Einstein field equations are then reduced to,
\begin{eqnarray}
{G^0}_0 = \mathrm{e}^{-C}\left(\frac{\dot{R}^2}{R^2} + \frac{\dot{A}\dot{R}}{R}\right) - \mathrm{e}^{-A}\left(2\frac{R''}{R} + \frac{R'^2}{R^2} - \frac{A'R'}{R}\right) + \frac{1}{R^2} = \kappa\rho, \label{etensor1}
\end{eqnarray}
\begin{eqnarray}
{G^1}_0 & = & \mathrm{e}^{-A}\left(2\frac{\dot{R'}}{R} - \frac{\dot{A}R'}{R} - \frac{\dot{R}C'}{R}\right) = 0, \label{etensor2} \\
{G^1}_1 & = & \mathrm{e}^{-C}\left(2\frac{\ddot{R}}{R} + \frac{\dot{R}^2}{R^2} - \frac{\dot{C}\dot{R}}{R}\right) - \mathrm{e}^{A}\left(\frac{R'^2}{R^2} + \frac{C'R'}{R}\right) + \frac{1}{R^2} = -\kappa p, \label{etensor3}
\end{eqnarray}
\begin{eqnarray}
{G^2}_2 = {G^3}_3 & = & \frac{1}{4}\mathrm{e}^{-C}\left(4\frac{{\ddot{R}}}{R}
- 2\frac{\dot{C}{\dot{R}}}{R} + 2\frac{\dot{A}\dot{R}}{R} +
2\ddot{A} + \dot{A}^2  - \dot{C}\dot{A}\right) -
\nonumber \\
& & - \frac{1}{4}\mathrm{e}^{-A}\left(4\frac{R''}{R} + 2\frac{C'R'}{R} -
2\frac{A'R'}{R} + 2C'' + C'^2 - C'A'\right) \nonumber \\
& = & -\kappa p,
\end{eqnarray}
where the dot and the prime mean partial derivatives with respect to $ t $ and $ r $, respectively. We also have $ c = G = 1 $ and $ \kappa = 8\pi $, a choice that turns the units into the so-called geometric units. In these units, the distances are given in $10^9$ pc, the time unit is in $3.26\times 10^9$ years and mass is expressed in units of $2.09\times 10^{22} M_{\odot}$. In order to solve the above system we assume the equation of state $ p = 0 $, i.e., evolution by gravitation only and the fluid will move along timelike geodesics. Differentiating equation (\ref{velocityfield}) we have, 
\begin{eqnarray}
\dot{u}^{\alpha} = -\frac{1}{2}\dot{C} \mathrm{e}^{-C/2}{\delta^{\alpha}}_0,
\end{eqnarray}
where we clearly see that if we assume zero acceleration, we must have $ \dot{C} = 0 $. 
If we use the following coordinate transformation,
\begin{equation}
\mathrm{d}\bar{t} = \mathrm{e}^{C/2}\mathrm{d}t, \label{coordtrans}
\end{equation}
we can find a value for the function $C(r,t)$. Using equation (\ref{coordtrans}), the new velocity field is given by,
\begin{equation}
{u}^{\bar{\alpha}} = \mathrm{e}^{C/2} {\delta^{\bar{\alpha}}}_0, \label{velocityfield3}
\end{equation}
or,
\begin{equation}
\mathrm{d}{x}^{\alpha} = \mathrm{e}^{C/2} \mathrm{d}\bar{t} {\delta^{\alpha}}_0,
\end{equation}
which can be written as,
\begin{equation}
\mathrm{d}{x}^{\alpha} = \mathrm{d}t {\delta^{\alpha}}_0.
\end{equation}
As a coordinate transformation cannot change the velocity field, equations (\ref{velocityfield}) and (\ref{velocityfield3}), imply that $ C = 0 $. Applying this condition in equation (\ref{etensor2}) leads to the result,
\begin{equation}
\mathrm{e}^{-A}\left(2\frac{\dot{R'}}{R} - \frac{\dot{A}R'}{R}\right) = 0.
\end{equation}
Simplifying by the product rule for derivatives (in $t$) we have,
\begin{equation}
\frac{\partial}{\partial t}(\mathrm{e}^{-A/2}R') = 0.\label{etensor2int}
\end{equation}
Now there are two particular solutions to be considered for the above equation. The one for $ R' = 0 $ is studied as a different case, the Datt-Ruban solution, where it is considered a fluid of charged dust, instead of having only a dust solution, and may lead to a $ neck $ or $ wormhole$ (see Plebanski \& Krasi\'nski 2006, p. 384). For the other case, with $ R' \neq 0 $, integrating equation (\ref{etensor2int}) in $ t $ we have,
\begin{equation}
\mathrm{e}^{A} = \frac{R'^2}{1 + 2E(r)}, \label{etensor2int2}
\end{equation}
where $ E(r) $ is an arbitrary function. The line element (\ref{lineel}) now takes the form,
\begin{equation}
\mathrm{d}S^2 = \mathrm{d}t^2 - \frac{R'^2}{1 + 2E(r)}\mathrm{d}r^2 - R^2(t,r)(\mathrm{d}\theta^2 + \sin^2{\theta} \mathrm{d}\phi^2).
\label{metricLT}
\end{equation}
Now assuming $ A \neq 0 $ and remembering that $C = 0$, we can substitute equation (\ref{etensor2int2}) into the field equation (\ref{etensor3}), multiply it by $ R^2\dot{R} $, and find,
\begin{equation}
2\ddot{R}\dot{R}R + \dot{R}^3 - (1 + 2E)\dot{R} + \dot{R} = 0.
\end{equation}
The above equation can be reduced using the product rule of derivatives, as follows,
\begin{equation}
\frac{\partial}{\partial t} (R\dot{R}^2 - 2ER) = 0.
\end{equation}
Integrating the above expression we have, 
\begin{equation}
	\dot{R}^2 = 2E(r) + 2\frac{M(r)}{R}, \label{fieldeq1}
\end{equation}
where $ M(r) $ is another arbitrary function. The equation (\ref{fieldeq1}) has the same form as the energy equation for a Newtonian radial motion in a Coulomb potential. The terms in the equation can be interpreted by using this analogy, with $ M(r) $ being the active gravitational mass, and $ E(r) $ playing the role of the total energy within the shell of radius $ r $. The same procedure of assuming $ A \neq 0 $ and having $C = 0$ applied to the field equation (\ref{etensor1}) will give,
\begin{equation}
\frac{\dot{R}^2}{R^2} + \frac{\dot{A}\dot{R}}{R} - \frac{1 + 2E(r)}{R'^2}\left(2\frac{R''}{R} + \frac{R'^2}{R^2} - \frac{A'R'}{R}\right) + \frac{1}{R^2} = \kappa\rho. \label{massdensity1}
\end{equation}
Considering equation (\ref{etensor2int2}), the derivatives of the function $A(r,t)$ with respect to $t$ and $r$ are, respectively,
\begin{eqnarray}
\left\{
\begin{array}{l}
\displaystyle \dot{A}\ = 2\frac{\dot{R'}}{R}, \\ \\
\displaystyle A' = 2\frac{R''}{R} + \frac{2E'}{1 + 2E}.  \label{derivA}
\end{array}
\right.
\end{eqnarray}
Substituting these derivatives into equation (\ref{massdensity1}) and using equation (\ref{fieldeq1}) we have
\begin{equation}
\frac{1 + 2E}{R^2} + \frac{2E'}{R'R} - \frac{2M'}{R'R^2} - \frac{1 + 2E}{R'^2}\left[\frac{R'^2}{R^2} - \frac{2R'E'}{R(1 + 2E)}\right] = \kappa\rho. \label{massdensity2}
\end{equation}
The first and fourth terms of the left-hand side cancel, as well as the second and fifth terms. The result is a constraint for the mass density in this geometry, given by the expression below,
\begin{equation}
	8 \pi \rho = \frac{2M'}{R^2R'}. \label{fieldeq2}
\end{equation}

The equations (\ref{metricLT}), (\ref{fieldeq1}), (\ref{fieldeq2}) define the Lema\^itre-Tolman-Bondi model. It can be seen that there are two singularities arising from these equations. One of them occurs when $ R = 0 $ and $ M'(r) \neq 0 $, and is interpreted as the Big Bang singularity. The other one comes where $ R' = 0 $ and $ M'(r) \neq 0 $, and it defines a shell crossing singularity, where the mass density diverges. This shell crossing separates two regions of the space-time with different velocities, and could indicate a breakdown in the assumptions of the model. However, these real singularities can be avoided by an appropriate choice of the functions $ M(r) $, $ E(r) $ (Hellaby \& Lake, 1984, 1985). 

The field equation (\ref{fieldeq1}) has three different solutions, depending on the range of the function $ E(r) $. These three possible distinct solutions are shown below. 
\\
For $E < 0$,
\begin{equation}
\left\{
\begin{array}{cc}
\displaystyle R(r,t) = - \frac{M}{2E}(1 - \cos{\eta}), \\ \\
\displaystyle \eta  - \sin{\eta} = \frac{-2E^{3/2}}{M}[t - t_B(r)]. \label{elipt}
\end{array}
\right.
\end{equation}
\\
For $E = 0$,
\begin{equation}
\displaystyle R(r,t) = \left\{\frac{9}{2}M(r)[t - t_B(r)]^2\right\}^{1/3}. \label{para}
\end{equation}
\\
For $E > 0$,
\begin{equation}
\left\{
\begin{array}{cc}
\displaystyle R(r,t) = \frac{M}{2E}(\cosh{\eta} - 1), \\ \\
\displaystyle \sinh{\eta} - \eta = \frac{(2E)^{3/2}}{M}[t - t_B(r)]. \label{hyper}
\end{array}
\right.
\end{equation}
These solutions were interpreted by Bondi (1947) as different systems. For $ E(r) < 0 $ and $ E(r) > 0 $ the system is bound and unbound, and these cases are known as elliptic and hyperbolic solutions, respectively. The simplest solution, the case $ E(r) = 0 $, is the parabolic solution and is interpreted by Bondi as a marginally bound system with no excess and no loss of energy. Another interpretation for this function is of a $ local $ curvature, dependent on $ r $. For every subspace of $ t  =$ constant, there is a different curvature, unlike the Friedmann models, where the curvature is a $ global $ characteristic of that spacetime. The picture is that the three distinct solutions can be considered as the same cosmological model, as they can simply hold in different regions of the spacetime. These equations contain a third arbitrary function, $ t_B(r) $. For $ t = t_B(r) $, one can cleary see that all three equations (\ref{elipt}), (\ref{para}) and (\ref{hyper}) lead to the same result, $ R(r,t) = 0 $. This value defines a singularity called the bang time, which in this geometry is position-dependent.

Another common notation of the LTB metric is the Bonnor's notation. In this case the line element and the field equations are given by,
\begin{equation}
\mathrm{d}S^2 = \mathrm{d}t^2 - \frac{R'^2}{f^2}\mathrm{d}r^2 - R^2(\mathrm{d}\theta^2 + \sin^2{\theta} \mathrm{d}\phi^2), \label{metricLTbonnor}
\end{equation}
\begin{equation}
	\frac{\dot{R}^2}{2} - \frac{F(r)}{4}\frac{1}{R}  = \frac{(f^2 - 1)}{2}, \label{fieldone}
\end{equation}
\begin{equation}
	8\pi\rho = \frac{F'}{2R'R^2}. \label{fieldtwo}
\end{equation}
We can see that the field equation (\ref{fieldone}) can be interpreted as an energy equation similarly as equation (\ref{fieldeq1}), where the first and second terms can be interpreted as an ``kinectic" and ``potential" energy terms, while the right-hand side is related to the ``total" energy of the system, basically the same interpretation as the one given by Bondi. The arbitrary functions in this notation are related to Bondi's as follows,
\begin{equation}
4M(r) = F(r), \label{massfunc}
\end{equation}
\begin{equation}
1 + 2E(r) = f^2(r),
\end{equation}
\begin{equation}
t_B(r) = \beta(r).
\end{equation}

We can see from the equation (\ref{fieldone}) that the functions related in equation (\ref{massfunc}) can be interpreted as the mass contained in a radius $ r$, a concept of mass defined as coordinate-dependent. We can now rewrite the solutions for $R(r,t)$ in terms of the Bonnor's notation, for the different intervals.
\\
For $f^2 < 1$,
\begin{equation}
\left\{
\begin{array}{cc}
\displaystyle R(r,t) = \frac{F}{4}\frac{(1 - 2\cos{2\Theta})}{| f^2 - 1 |}, \\ \\
\displaystyle t + \beta = \frac{F}{4}\frac{(2\Theta  - \sin{2\Theta})}{| f^2 - 1 |^{-3/2}}. \label{eliptB}
\end{array}
\right.
\end{equation}
\\
For $f^2 = 1$,
\begin{equation}
\displaystyle R(r,t) = \frac{(9F)^{1/3}}{2}[t + \beta]^{2/3}. \label{paraB}
\end{equation}
\\
For $f^2 > 1$,
\begin{equation}
\left\{
\begin{array}{cc}
\displaystyle R(r,t) = \frac{F}{4}\frac{(2\cosh{2\Theta} - 1)}{(f^2 - 1)}, \\ \\
\displaystyle t + \beta = \frac{F}{4}\frac{(\sinh{2\Theta} - 2\Theta)}{(f^2 - 1)^{-3/2}}. \label{hyperB}
\end{array}
\right.
\end{equation}
The Bonnor's notation will be used from now on for its simplicity.

\section{Observational Quantities}\label{observational}

Among the most relevant results that can be obtained in a cosmological model are relations between observables and the functions that define the model itself, as one of the main goals of a cosmological model is their ability to predict and describe many properties of the observable universe. The next subsections will discuss some useful quantities for studying cosmology. In section 1.2.1 we derive the commonly used distances, in section 1.2.2 we obtain the expressions for the number counts and the global densities and, finally, in the sections 1.2.3 and 1.2.4 we derive two concordant methods for finding the redshift.

\subsection{Distances}

With the results obtained in the previous section we can derive expressions for relevant quantities as a function of the $ redshift $. For that, first we shall define the distances to be used in these equations. 

\begin{figure}[!t]
\centering
\includegraphics[scale=1]{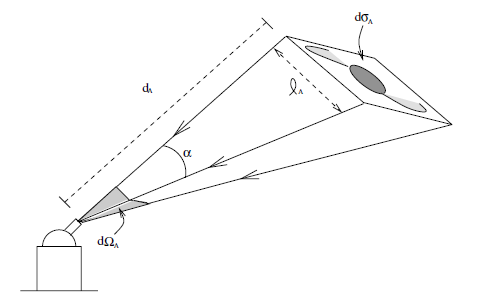}
\caption[The observer area distance scheme]{The area distance $d_A$ is obtained by the relationship between the
intrinsic cross-sectional area $\mathrm{d}\sigma_A$ of the source, measured at its restframe,
and the solid angle $\mathrm{d}\Omega_A$ measured by the observer (Ribeiro 2005).}
\label{angdist}
\end{figure}

According to Ellis (1971), the `observer area distance',  or simply `area distance' $ d_A $ is defined as below,
\begin{equation}
(d_A)^2 = \frac{\mathrm{d}\sigma_A}{\mathrm{d}\Omega_A},
\end{equation}
where $ \mathrm{d}\sigma_A $ is the intrinsic cross-sectional area of the source and $ \mathrm{d}\Omega_A $ is the solid angle measured by the observer, over a bundle of null geodesics. The expression for the area distance in the LTB geometry is
\begin{equation}
(\mathrm{d}_A)^2 = R^2 \frac{\mathrm{d}\theta \sin{\theta}\mathrm{d}\phi}{\mathrm{d}\theta\sin{\theta}\mathrm{d}\phi} = R^2,
\end{equation}
which is valid if we consider that the light travels from the source to the observer through null geodesics. This distance is also known in the literature as `angular diameter distance'. We can see in Figure \ref{angdist} how the area distance is defined geometrically.

\begin{figure}[!t]
\centering
\includegraphics[scale=1]{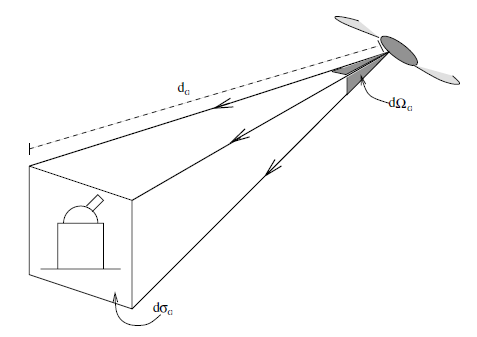}
\caption[The galaxy area distance scheme]{The galaxy area distance $d_G$ is obtained by the relationship between
the cross-sectional area $d_{\sigma G}$ measured at the observer's restframe,
and the solid angle $\mathrm{d}\Omega_G$ measured at the source's rest-frame (Ribeiro 2005).}
\label{galdist}
\end{figure}

The luminosity distance $ d_L $ is defined by the expression,
\begin{equation}
\mathcal{F} = \frac{L}{4\pi (d_L)^2},
\end{equation}
where $ \mathcal{F} $ is the radiation flux measured by the observer, and $ L $ is the intrinsic luminosity of the object. These distances are connected by the famous Etherington (1933) reciprocity theorem (Ellis 1971, 2007),
\begin{equation}
d_L = d_A (1 + z)^2 = d_G(1 + z),
\end{equation}
which is valid for any space-time, given that the metric that describes this space-time has symmetric connections and under the assumption that light travels along null geodesics. Here $ z $ is the redshift and $ d_G $ is the `galaxy area distance', or `angular size distance', which is a distance measured from the point-source, the opposite of the area distance. For the LTB space-time, this theorem states that,
\begin{equation}
d_L = R(1 + z)^2. \label{dLz}
\end{equation}
Figure \ref{galdist} shows a diagram for the galaxy area distance illustrating how it can be obtained geometrically.

\subsection{Number Counts and Densities}

The cumulative number counts $ N $ of sources is given by the following general differential expression (Ellis 1971),
\begin{equation}
	dN = (d_A)^2d\Omega_0[n(-k^{\alpha} u_{\alpha})]_P\mathrm{d}\lambda, \label{dnumbercount}
\end{equation}
\begin{figure}[!h]
\centering
\includegraphics[scale=1]{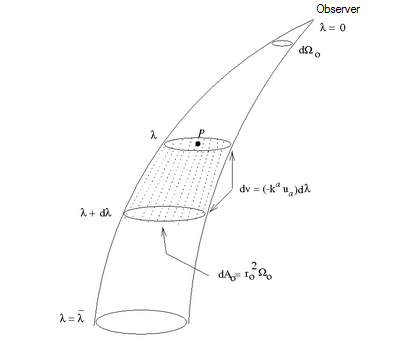}
\caption{A section of a bundle of null geodesics which subtends a solid angle $\mathrm{d}\Omega_0$ at the observer's position $\lambda = 0$ (adapted from Moura Jr. 1997).}
\label{cone}
\end{figure} 
where $ n $ is the numerical density per proper volume in the solid angle $ \mathrm{d}\Omega_0 $ over a displacement of the affine parameter $\mathrm{d}\lambda$, where $k^{\alpha}$ is the radiation flux vector and $u^{\alpha}$ is the four velocity of the observer. Figure \ref{cone} illustrates how this expression is obtained. To proceed now we need to reduce the above equation in terms of the LTB spacetime. For that, we initially project the tangent vector of the null geodesic $k^{\alpha} = \mathrm{d}x^{\alpha}/\mathrm{d}\lambda$ in the observer's spacelike hypersurface (Moura Jr. 1997) by using the projection tensor defined by,
\begin{equation}
{h^{\alpha}}_{\beta} = {\delta^{\alpha}}_{\beta} + u^{\alpha}u_{\alpha}, \label{projecttensor}
\end{equation}
which results in,
\begin{equation}
{h^{\alpha}}_{\beta}k^{\beta} = k^{\alpha} - u^{\alpha}u_{\beta}k^{\beta}. \label{project2}
\end{equation}
\begin{figure}[!h]
\centering
\includegraphics[scale=1]{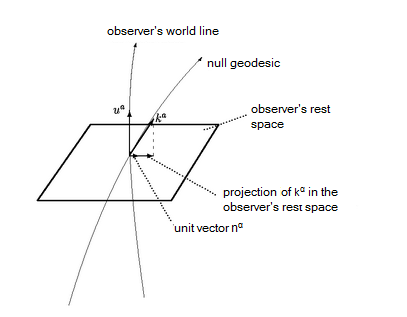}
\caption[Diagram for the directions of the vector and its projection in the observer's rest space]{The projection of the vector $k^{\alpha}$ on the observer's rest space, where $n^{\alpha}$ is the unit vector in the direction of this projection and it is ortogonal to the observer's four-velocity $u^{\alpha}$ (adpated from Moura Jr. 1997).}
\label{proj}
\end{figure}	
We can see in Figure \ref{proj} that the projection of $k^{\alpha}$ is perpendicular to the velocity of the observer $u^{\alpha}$, or that,
\begin{equation}
n_{\alpha}u^{\alpha} = 0, \label{prodint}
\end{equation}
where $n^{\alpha}$ is the unit vector in the direction of the projection of $k^{\alpha}$. As $n^{\alpha}$ is a space-like vector, its normalization is given by, 
\begin{equation}
n^{\alpha}n_{\alpha} = -1, \label{normaln}
\end{equation}
and
\begin{equation}
u^{\alpha}u_{\alpha} = 1. \label{normalu}
\end{equation}
Considering equation (\ref{project2}), we can write now that,
\begin{equation}
k^{\alpha} = Ln^{\alpha} + u^{\alpha}u_{\beta}k^{\beta}, \label{modproj}
\end{equation}
where $L$ is the modulus of the spacelike projection to be determined. From the above equation we have,
\begin{equation}
k^{\alpha}k_{\alpha} = (Ln^{\alpha} + u^{\alpha}u_{\beta}k^{\beta})(Ln_{\alpha} + u_{\alpha}u_{\beta}k^{\beta}),
\end{equation}
or, if we multiply the vectors,
\begin{equation}
L^2n^{\alpha}n_{\alpha} + (k^{\beta}u_{\beta})^2(u^{\alpha}u_{\alpha}) + Ln^{\alpha}k^{\beta}u_{\beta}u_{\alpha} + Ln_{\alpha}k^{\beta}u_{\beta}u^{\alpha} = 0.
\end{equation}
In view of equation (\ref{prodint}), the expression above reduces to,
\begin{equation}
L^2n^{\alpha}n_{\alpha} + (k^{\beta}u_{\beta})^2(u^{\alpha}u_{\alpha}) = 0.
\end{equation}
Remembering equations (\ref{normaln}) and (\ref{normalu}), we end up with the following expression,
\begin{equation}
L^2 - (k^{\beta}u_{\beta})^2 = 0,
\end{equation}
 or,
\begin{equation}
L = \pm k^{\beta}u_{\beta}.
\end{equation}
If we choose the minus sign, for an incoming light ray, we have that,
\begin{equation}
L = -k^{\beta}u_{\beta}. \label{modulus}
\end{equation}
Substituting back into equation (\ref{modproj}), we have that,
\begin{equation}
k^{\alpha} = -(k^{\beta}u_{\beta})(n^{\alpha} - u^{\alpha}).
\end{equation}
The above equation is simply a decomposition of the vector $k^{\alpha}$ in two parts, one in the rest space of the observer and another in the observer's world line. As $k^{\alpha} = \mathrm{d}x^{\alpha}/\mathrm{d}\lambda$, it follows that,
\begin{equation}
\mathrm{d}x^{\alpha} = -(k^{\beta}u_{\beta})(n^{\alpha} - u^{\alpha})\mathrm{d}\lambda,
\end{equation}
and then, if we have an increase in the affine parameter $\lambda$, an increase in the quantity $(k^{\beta}u_{\beta})\mathrm{d}\lambda$ will follow, and the spatial displacement will be,
\begin{equation}
\mathrm{d}v = -(k^{\beta}u_{\beta})\mathrm{d}\lambda.
\end{equation}
Now if we substitute the vector $u^{\alpha} = {\delta^{\alpha}}_ 0$ in the above equation, we find that,
\begin{equation}
\mathrm{d}v = -(k^0)\mathrm{d}\lambda,
\end{equation}
and, as $k^0 = \mathrm{d}x^0/\mathrm{d}\lambda = \mathrm{d}t/\mathrm{d}\lambda$, we have that,
\begin{equation}
\mathrm{d}v = -\mathrm{d}t. \label{displacement}
\end{equation}
Since we are dealing with a radial past null geodesic (incoming light ray), we have $ \mathrm{d}S^2 = 0 = \mathrm{d}\theta^2 = \mathrm{d}\phi^2 $ in equation (\ref{metricLTbonnor}), or
\begin{equation}
	\frac{dt}{d\lambda} = - \left(\frac{R'}{f}\right)\left(\frac{\mathrm{d}r}{\mathrm{d}\lambda}\right). \label{geodesicLT}
\end{equation}
Substituting equation (\ref{geodesicLT}) in equation (\ref{displacement}), the displacement is finally given by,
\begin{equation}
\mathrm{d}v = \frac{R'}{f}dr.
\end{equation}
As we have the spherical symmetry of the LTB geometry, the equation (\ref{dnumbercount}) reduces to,
\begin{equation}
	\mathrm{d}N = 4\pi n{R}^2\frac{R'}{f}\mathrm{d}r. \label{numbersources}
\end{equation}
Considering the local density equation (\ref{fieldtwo}), we can find an expression for the quantity $ n $,
\begin{equation}
	n = \frac{\rho}{\mathcal{M}_g} = \frac{F'}{16\pi \mathcal{M}_g{R}'{R}^2}, \label{nvol}
\end{equation}
where $ \mathcal{M}_g $ is the $rest$ mass of the cosmological sources, usually galaxies. We assume that there is no evolution in mass of the sources. We are also assuming that the galaxies have negligible relative velocities, otherwise we would have a non-zero pressure, which contradicts the initial assumption of zero pressure for our LTB spacetime. Substituting equation (\ref{nvol}) into the expression (\ref{numbersources}), we find,
\begin{equation}
	N(r) = \frac{1}{4\mathcal{M}_g}\int_C \frac{F'}{f}dr. \label{numberLT}
\end{equation}

The global volume and the $global$ density, which is different from the previously defined $local$ density (\ref{fieldtwo}), can be defined respectively as,
\begin{equation}
	V_i(r) = \frac{4}{3}\pi (d_i)^3, \label{volumegen} 
\end{equation}
\begin{equation}
	{\gamma^{*}}_i(r) = \frac{\mathcal{M}_gN(r)}{V_i(r)}, \label{densityLTgen}
\end{equation}
where the indexes $i = \{L, A, G\}$ are related to the different cosmological distances described in the previous section. For example, if we take the luminosity distance, the equations above reduce to
\begin{equation}
	V_L(r) = \frac{4}{3}\pi (d_L)^3 = \frac{4}{3}\pi {R}^3(1 + z)^6, \label{volumeLT}  
\end{equation}
\begin{equation}
	{\gamma^{*}}_L(r) = \frac{\mathcal{M}_gN(r)}{V_L(r)} = \frac{3}{16\pi {R}^3(1 + z)^6} \int_C \frac{F'}{f}\mathrm{d}r. \label{densityLT}
\end{equation}

\subsection{Redshift}

Now we are able to find an expression for the fundamental quantity known as the redshift in the LTB model. The most general and geometrically independent expression for the redshift is given by Ellis (1971) as follows,
\begin{equation}
	1 + z = \frac{(u^{\alpha}k_{\alpha})_{source}}{(u^{\beta}k_{\beta})_{observer}}. \label{redshift1}
\end{equation}
In comoving and synchronous spherically symmetric coordinates, the expression above becomes,
\begin{equation}
	1 + z = \frac{\mathrm{d}t}{\mathrm{d}\lambda}\Big|_{\lambda=\bar{\lambda}}\left(\frac{\mathrm{d}t}{\mathrm{d}\lambda}\Big|_{\lambda=0}\right)^{-1}, \label{redshift2}
\end{equation}
where $ \bar{\lambda} $ is an arbitrary value for the affine parameter $\lambda$ along the geodesic and $\lambda = 0$ is what we call ``here and now", the observer's position. 

In order to solve the equation above we shall need first to obtain the so-called regularity conditions. The solutions of the field equation (\ref{fieldone}), given by the expressions (\ref{eliptB}), (\ref{paraB}) and (\ref{hyperB}), have to be locally Euclidean in the vicinity of the origin ($r \approx 0$). We can obtain these regularity conditions at the origin by following a simple procedure (Ribeiro 1993). Consider a displacement in the 2-surface $ t = $ constant, $\phi = $ constant in equation (\ref{metricLTbonnor}),
\begin{equation}
-\mathrm{d}S^2 = \frac{R'^2}{f^2}\left(\mathrm{d}r^2 + \frac{R^2f^2}{R'^2}\mathrm{d}\theta^2\right).
\end{equation}
For this 2-surface to be Euclidean at $ r = 0$, we need that $R'^2f^{-2} \rightarrow $ constant ($\neq 0$), and then
\begin{equation}
\frac{f^2R^2}{R'^2} \sim r ^2,
\end{equation}
or, as in Bonnor (1974),
\begin{equation}
\displaystyle \lim_{r \to 0} \frac{R'^2r^2}{f^2R^2} = 1.
\end{equation}
Supposing now that $ f$ is approximately constant for small $r$, the equation 
\begin{equation}
\frac{R'^2}{R^2} = \frac{f^2}{r^2}
\end{equation}
can be integrated to get $R = r^f$, and therefore, 
\begin{equation}
\displaystyle \lim_{r \to 0} \frac{R'}{f} = \lim_{r \to 0} r^{f - 1}. \label {limcond}
\end{equation}
If $ f < 1 $, the limit goes to infinity and if $ f > 1 $ it is zero. Hence, it can only converge to a nonzero constant if $ f = 1$, and $R \sim r$ as $ r \rightarrow 0 $. Now if we also assume that $ F \approx 0$ as $r \rightarrow 0$, the field equation (\ref{fieldone}) necessarily reduces to $\dot{R} = 0$, and, as we already have seen that $R \approx r$ when $r \rightarrow 0$, we have that $R'(0) = 1$. The regularity conditions at $ r \approx 0$ are then,
\begin{equation}
\left\{
\begin{array}{cc}
	R = r, \\ f = 1, 
\end{array}
\right.
\\
\begin{array}{cc}
	R' = 1, \\ F = 0. \label{conditionsreg}
\end{array}
\end{equation}
Taking into consideration these conditions in equation (\ref{geodesicLT}), and substituting the latter into equation (\ref{redshift2}), we find
\begin{equation}
	1 + z = \left[\frac{R'}{f}\frac{\mathrm{d}r}{\mathrm{d}\lambda}\right]_{\lambda = \bar{\lambda}}\left(\frac{\mathrm{d}r}{\mathrm{d}\lambda}\Big|_{\lambda = 0}\right)^{-1}. \label{redshiftgeneral}
\end{equation}

We now will use the Lagrangean method to solve the right-hand side of the differential equation above (Ribeiro 1992a). The Lagrangean for the LTB metric with $ \mathrm{d}\theta^2 = \mathrm{d}\phi^2 = 0 $ and its two Euler-Lagrange equations are, respectively,
\begin{equation}
	\mathrm{L} = \left(\frac{\mathrm{d}S}{\mathrm{d}\lambda}\right)^2 = \left(\frac{\mathrm{d}t}{\mathrm{d}\lambda}\right)^2 - \left(\frac{R'}{f}\frac{\mathrm{d}r}{\mathrm{d}\lambda}\right)^2, \label{lag}
\end{equation}
where the Lagrange equations of the second kind are written as,
\begin{equation}
\frac{\mathrm{d}}{\mathrm{d}\lambda}\frac{\partial \mathrm{L}}{\partial (\mathrm{d}x^{\nu}/\mathrm{d}\lambda)} - \frac{\partial \mathrm{L}}{\partial x^{\nu}} = 0, \label{lageq2}
\end{equation}
which in the radial case turn to,
\begin{equation}
	\displaystyle \frac{\mathrm{d}^2t}{\mathrm{d}\lambda^2} + \left(\frac{\mathrm{d}r}{\mathrm{d}\lambda}\right)^2 \frac{R'\dot{R}}{f^2} = 0, \label{lagrange1}
\end{equation}
\begin{equation}
	\displaystyle \frac{\mathrm{d}^2r}{\mathrm{d}\lambda^2} + \frac{1}{R'}\left(\frac{\mathrm{d}r}{\mathrm{d}\lambda}\right)^2\left(R'' - \frac{f'R'}{f}\right) + 2 \ \frac{\mathrm{d}r}{\mathrm{d}\lambda}\frac{\mathrm{d}t}{\mathrm{d}\lambda}\frac{\dot{R'}}{R'} = 0. \label{lagrange2}
\end{equation}
Substituting the past radial null geodesic equation (\ref{geodesicLT}) into equations (\ref{lagrange1}) and (\ref{lagrange2}), we have,
\begin{equation}
	\displaystyle \displaystyle \frac{\mathrm{d}^2t}{\mathrm{d}\lambda^2} - \left(\frac{\mathrm{d}t}{\mathrm{d}\lambda}\right)^2 \frac{\dot{R}}{R'} = 0,
\end{equation}
\begin{equation}
	\displaystyle \frac{\mathrm{d}^2r}{\mathrm{d}\lambda^2} + \frac{1}{R'}\left(\frac{\mathrm{d}r}{\mathrm{d}\lambda}\right)^2\left(R'' - \frac{f'R'}{f}\right) - 2 \ \frac{\mathrm{d}r}{\mathrm{d}\lambda}\frac{\mathrm{d}r}{\mathrm{d}\lambda}\frac{R'\dot{R'}}{f} = 0.
\end{equation}
Integrating the above equations once, we find,
\begin{equation}
	\displaystyle \frac{\mathrm{d}t}{\mathrm{d}\lambda} = \frac{1}{I + C_1}, \label{lag1int}
\end{equation}
\begin{equation}
	\displaystyle \frac{\mathrm{d}r}{\mathrm{d}\lambda} = \left[\int\left(\frac{R''}{R'} - \frac{f'}{f} - \frac{2\dot{R'}}{f}\right) \mathrm{d}\lambda + C_2\right]^{-1}, \label{lag2int}
\end{equation}
where
\begin{equation}
	I \equiv \int\frac{\dot{R'}}{R'} \mathrm{d}\lambda, \label{int_I}
\end{equation}
and $C_1$ and $C_2$ are two constants of integration that can be deduced if we substitute the equations (\ref{lag1int}) e (\ref{lag2int}) back into the geodesic equation (\ref{geodesicLT}). Then we find,
\begin{equation}
	\int\left(\frac{R''}{R'} - \frac{f'}{f} - \frac{2\dot{R'}}{f}\right) \mathrm{d}\lambda + C_2 = - \frac{R'}{f}(I + C_1), \label{geodconst}
\end{equation}
which is valid for any value of $ \lambda $. Assuming now we have that $\lambda = 0$ at $r = 0$ and using the regularity conditions (\ref{conditionsreg}), we conclude that
\begin{equation}
	C_1 = - C_2. \label{const}
\end{equation} 
As $C_1$ and $C_2$ are arbitrary constants, we choose now $ C_1 = -1 $ and $ C_2 = 1 $, as we are dealing with the past null geodesic. Therefore,
\begin{equation}
	\frac{\mathrm{d}t}{\mathrm{d}\lambda} = \frac{1}{I - 1}, \label{finallag1}
\end{equation}
\begin{equation}
	\frac{\mathrm{d}r}{\mathrm{d}\lambda} = \frac{f}{(1 - I)R'}. \label{finallag2}
\end{equation}
The above choice of constants and regularity conditions show that equation (\ref{finallag2}) reduces to,
\begin{equation}
	\frac{\mathrm{d}r}{\mathrm{d}\lambda}\Big|_{\lambda = 0} = 1. \label{regcond}
\end{equation}
Substituting the above expression into equation (\ref{redshiftgeneral}), the final result is,
\begin{equation}
	z = \frac{I}{1 - I}. \label{redshiftI}
\end{equation}
The integral (\ref{int_I}) can be finally written in terms of the following differential equation,
\begin{equation}
	\frac{\mathrm{d}I}{\mathrm{d}r} = \frac{(1 - I)}{f}\dot{R}'. \label{der_I}
\end{equation}
Solutions for the differential equation (\ref{der_I}) as a function of $ r $ can be used to obtain a final expression for the redshift in terms of $ r $,  depending on the form of the functions $ f(r) $ e $ \dot{R'}(r,t) $. This will be done with the aid of the single null geodesic, in chapter 3.

\subsection{Alternative Method for the Redshift}

There is another way to obtain an expression for the redshift. We present here the procedure showed by Pleba\'nski \& Krasi\'nski (2006), originally done by Bondi (1947). Let two light rays be emitted in the same direction, the second one being emitted after a small time-interval $\Delta T$. The equation for the first and the second rays are, respectively, 
\begin{equation}
t_1 = T(r), \ \ 
t_2 = T(r) + \Delta T(r).
\end{equation}
Both rays must obey the past null geodesic equation (\ref{geodesicLT}), so
\begin{eqnarray}
&  &\frac{\mathrm{d}T}{\mathrm{d}r} = - \frac{R'[T(r), r]}{f}, \label{geodesicT} \\
&  &\frac{\mathrm{d}(T + \Delta T)}{\mathrm{d}r} = - \frac{R'[T(r) + \Delta T, r]}{f}. \label{geodesicTtau}
\end{eqnarray}
As $ \Delta T(r) $ is assumed to be small, we have to first order in $ \Delta T $,
\begin{equation}
R'[T(r) + \Delta T, r] = R'[T(r), r] + \Delta T(r)\dot{R}'[T(r), r]. \label{aproxtau}
\end{equation}
Substituting the above expression into equation (\ref{geodesicTtau}) and using equation (\ref{geodesicT}) we have, 
\begin{equation} 
\frac{\mathrm{d}T}{\mathrm{d}r} + \frac{\mathrm{d}\Delta T}{\mathrm{d}r} = - \frac{R'[T(r), r]}{f} - \Delta T(r)\frac{\dot{R}'[T(r), r]}{f}.
\end{equation}
According to equation (\ref{geodesicT}), the first terms of the left-hand and right-hand sides cancel out and the result is,
\begin{equation}
\frac{\mathrm{d}\Delta T}{\mathrm{d}r} = - \Delta T(r)\frac{\dot{R}'[T(r), r]}{f}. \label{geodesictau}
\end{equation}
As stated by equation (\ref{redshift1}), the redshift can be defined as a relation between the period $\Delta T $ of emission of the wave and the period measured at the observer's position,
\begin{equation} 
\frac{\Delta T(r_{source})}{\Delta T(r_{obs})} = 1 + z(r_{source}).
\end{equation}
Now we keep the observer at a fixed position and consider the source at two positions, $ r_{source}$ and $(r_{source} + dr)$. Differentiating the above equation with respect to $r$ results in,
\begin{equation}
\frac{\mathrm{d}\Delta T_{source}}{\mathrm{d}r} - \frac{\mathrm{d}z}{\mathrm{d}r}\Delta T_{obs} = (1 + z) \frac{\mathrm{d}\Delta T_{obs}}{\mathrm{d}r}.
\end{equation}
If we consider that at the source's position $\mathrm{d}\Delta T_{source}/\mathrm{d}r \sim 0 $, the expression above reduces to,
\begin{equation}
(1+z)\frac{\mathrm{d}\Delta T}{\mathrm{d}r} = - \Delta T \frac{\mathrm{d}z}{\mathrm{d}r}.
\end{equation}
where $\Delta T_{obs} = \Delta T$. Substituting this expression into equation (\ref{geodesictau}), we find
\begin{equation}
\frac{1}{1 + z} \frac{\mathrm{d}z}{\mathrm{d}r} = \frac{\dot{R}'[T(r), r]}{f}, \label{redshiftr}
\end{equation}
whose integration produces the following expression,
\begin{equation}
\ln{[1 + z(r)]} = \int_{r_{em}}^{r_{obs}} \frac{\dot{R}'[T(r), r]}{f} \mathrm{d}r.
\end{equation}

This result is the same as the one obtained by the Lagrangean method shown in the previous section. To see that, let us differentiate equation (\ref{redshiftI}) with respect to $r$, 
\begin{equation}
\frac{\mathrm{d}z}{\mathrm{d}r} = \frac{\mathrm{d}I}{\mathrm{d}r}\frac{1}{(1 - I)^2}. \label{der_zI}
\end{equation}
Now if we substitute equation (\ref{der_I}) into equation (\ref{redshiftr}), to eliminate the arbitrary functions $\dot{R}'$ and $f$, we find the relation,
\begin{equation}
\frac{1}{1 + z} \frac{\mathrm{d}z}{\mathrm{d}r} = \frac{\mathrm{d}I}{\mathrm{d}r}\frac{1}{1 - I},
\end{equation}
which is the same expression as equation (\ref{der_zI}), given that $(1 + z)^{-1} = (1 - I)$, by equation (\ref{redshiftI}).

\section{Particular Cases}\label{cases}

In order to find special cases of the LTB cosmology, we have to specify at least one of the arbitrary functions of the model. In this section, we will show how different specifications lead to different models. 

\subsection{Friedmann Models}

In order to obtain the Friedmann metric from the LTB one, it is necessary to assume that (Ribeiro 1992a, 1994),
\begin{equation}
R(r,t) = a(t) g(r), \ \  f(r) = g'(r). \label{functionsFried}
\end{equation}
Substituting these functions into equation (\ref{metricLTbonnor}) we have that,
\begin{equation}
\mathrm{d}S^2 = \mathrm{d}t^2 - a^2(t)\left\{\mathrm{d}r^2 + g^2(r)[\mathrm{d}\theta^2 + \sin \theta^2\mathrm{d}\phi^2]\right\},
\end{equation}
which is a Friedmann metric if
\begin{equation}
g(r) = \left\{
\begin{array}{cc}
\sin r, \\r, \\ \sinh r.
\end{array}
\right.
\end{equation}
Substituting the equations (\ref{functionsFried}) in equation (\ref{fieldtwo}), and integrating it, we have that,
\begin{equation}
\displaystyle \frac{F}{4} = \frac{4\pi}{3}\rho a^3 g^3. \label{densityFried}
\end{equation}
Now if we substitute equations (\ref{functionsFried}) and (\ref{densityFried}) into the field equation (\ref{fieldone}) we obtain the following result,
\begin{equation}
\dot{a}^2 = \frac{8 \pi}{3} \rho a^2 - K, \label{Friedeq}
\end{equation}
where $K = (1 -g')^2/g^2$. It is clear that $K =+1, 0, -1$ if $g = \sin r, r, \sinh r$, respectively, and that equation (\ref{Friedeq}) is the usual Friedmann equation, which can be written in the following form,
\begin{equation}
\frac{\dot{a}^2g^2}{2} - \frac{w}{ag} = - (1 - g'^2), \label{Friedeq2}
\end{equation}
where
\begin{equation}
w(r) = \frac{4 \pi}{3} \rho a^3g^3. \label{mass}
\end{equation}
Once again we can interpretate the Friedmann field equation (\ref{Friedeq2}) as an energy equation, as we did for the LTB model, with $4w(r) = F(r)$ acting as the gravitational mass inside the coordinate $r$. The right-hand side of the equation (\ref{Friedeq2}) can also be interpreted as the total energy of the system.

The function $\beta(r)$ gives the bang time of the model, and if we consider  $t = 0$, and $\beta(r) = 0$, then the hypersurface $ t = 0 $ is singular, that is, $ R = 0 $ everywhere, so $\beta(r)$ gives the age of the universe which in the LTB spacetime is different for observers at different radial coordinates $r$. In the Friedmann models this characteristic does not exist - the universe has the same age for all observers at any positions on a hypersurface of constant $t$, $i.e.$, the big bang is simultaneous. In the LTB spacetime, the big bang may have ocurred at different proper times in different locations. So, to reduce from the LTB metric to Friedmann we have to assume $\beta =$constant. We can link $\beta$ and the Hubble constant $H$ as follows,
\begin{equation}
\frac{\dot{R}}{R} = \frac{\dot{a}}{a} = H(t), \ \ \mathrm{for} \ \ \beta = \beta_0, \label{hubblebeta}
\end{equation}
where $\beta_0$ is a constant. If we consider the parabolic model given by the equation (\ref{paraB}), we can see that 
\begin{equation}
\beta_0 = \frac{2}{3H_0},
\end{equation}
where $H_0 = H(0) $. The above equation relates $\beta_0$ and the Hubble constant $H_0$ in an Einstein-de Sitter universe.

In a similar way, as discussed by Pleba\'nski \& Krasi\'nski (2006), the general Friedmann models can also be obtained from the LTB spacetime by the following choice of the functions:
\begin{eqnarray}
F(r) = 4 m_0 r^3; \ \ 
f^2(r) = 1 - kr^2; \ \ 
\beta(r) = \mathrm{constant}, \label{functionsfried}
\end{eqnarray}
where $ m_0 $ and $k$ are two constants, with $m_0$ acting as a mass component and $k$ being the curvature of the Friedmann spacetime. The last function, $ \beta(r)$, is considered constant or, more precisely, simultaneous for every point of the spacetime. This is a coordinate-dependent limit, but there is an invariant condition for this case, which is $\rho' = 0$. In each interval of $r$  in which $F'$ does not change sign, $F(r)$ can be used as independent variable instead of $r$, as equations (\ref{metricLTbonnor}) and (\ref{fieldtwo}) are covariant under the coordinate transformations $r = h(r')$. Applying this condition to equation (\ref{fieldtwo}), we have,
\begin{equation}
\displaystyle \rho = \frac{3}{16\pi}\left[\frac{\mathrm{d}(R^3)}{\mathrm{d}F}\right]^{-1} \ \Leftrightarrow \  R^3 - R^3(F_0) = \int_{F_0}^F \frac{3}{16 \pi \rho(\tilde{F})} d\tilde{F}. \label{rho}
\end{equation}
Then the condition of spatial homogeneity $\rho ' = 0$ implies that,
\begin{equation}
\frac{\mathrm{d}\rho}{\mathrm{d}F} = 0, \ \ \mathrm{or} \ \ \frac{\mathrm{d^2}(R^3)}{\mathrm{d}F^2} = 0. 
\end{equation}
If we use equations (\ref{eliptB}) and (\ref{hyperB}), we can solve the condition above for the different intervals of $f(r)$, resulting in,
\begin{equation}
\frac{1 - f^2}{F^{2/3}} = \mathrm{constant}, \ \ \ \ \beta(r) = \mathrm{constant}. \label{invariantfried}
\end{equation} 
The equations (\ref{invariantfried}) define the Friedmann limit invariantly. 

Now if we substitute the functions (\ref{functionsfried}) into the field equation (\ref{fieldone}), it takes the form,
\begin{equation}
\dot{R}^2 + kr^2 - \frac{2m_0r^3}{R} = 0.
\end{equation}
The field equation can be written in another way if we multiply it by $1/R^2$,
\begin{equation}
\frac{\dot{R}^2}{R^2} + \frac{kr^2}{R^2} - \frac{2m_0r^3}{R^3} = 0. \label{fried}
\end{equation}
The equation (\ref{fried}) can still be rewritten in the parameter form,
\begin{equation}
H + \Omega_k - \Omega_m = 0, \label{Friedparam}
\end{equation}
where $H$ is called the Hubble parameter, $\Omega_k$ is the curvature parameter and $\Omega_m$ is the mass parameter. These parameters are often used as estimates of the characteristics of the Universe, as the Friedmann model is considered the actual standard one. The current values for these parameters are $\Omega_k \simeq 0$, $\Omega_m \simeq 0.27$, with $H_0 \simeq 70$ km/s/Mpc (Komatsu et al.\ 2011). Here we are not considering the density of the vaccuum parameter $\Omega_{\Lambda}$, often associated to a cosmological constant $\Lambda$, commonly interpreted as a dark energy component responsible for an accelarated universe. The value of this component in the current standard Friedmann model is $\Omega_{\Lambda} \simeq 0.73$, and if we consider this parameter in the original field equation (\ref{einstein_eq}), it is written as,
\begin{equation}
G_{\mu \nu} = R_{\mu \nu} - \frac{1}{2}g_{\mu \nu}\mathrm{R} + \Lambda g_{\mu \nu}, \label{einstein_eqlambda}
\end{equation}
and the expression (\ref{Friedparam}) becomes,
\begin{equation}
H + \Omega_k - \Omega_m  - \Omega_{\Lambda} = 0, \label{Friedparam2}
\end{equation}
where $\Omega_{\Lambda} = (1/3)\Lambda R^2$. The equation for the mass density (\ref{fieldtwo}) is then given as,
\begin{equation}
\rho = \frac{3m_0r^2}{4\pi R'R^2}. \label{densityFRW}
\end{equation}
Using equation (\ref{densityFRW}) in equation (\ref{fried}), we can define a $critical$ $density$, given by
\begin{equation}
\rho_{cr} = \frac{3H^2R'R}{8\pi r}, \label{critdens}
\end{equation}
which defines the curvature of the universe in this model. If $\rho > \rho_{cr}$, then the universe is closed ($k > 0$); if $\rho < \rho_{cr}$, then the universe is open ($k < 0$), and if $\rho = \rho_{cr}$, the universe is flat ($k = 0$). The matter observed by telescopes accounts only for, approximately, $\rho_o \leq 0.2\rho_{cr}$, considering the actual value for $H_0$, which yields an estimate age for the Universe of $\sim 14$ Gyr.
The next section will explore the Einstein-de Sitter model, which assumes a flat universe ($k = 0$) and a simultaneous big bang.

\subsection{Einstein-de Sitter Model}

\indent Among many cosmologies that can be obtained from the LTB spacetime, a straighforward one is the Einstein-de Sitter (EdS) model. The EdS geometry is a case limit of the spherically symmetric matter dominated Friedmann cosmology with null pressure, where the Universe has the exactly necessary amout of energy to escape the collapse due to gravity. To generate it, the LTB functions take the form (Ribeiro 1992b),
\begin{equation}
	F(r) =  \frac{8}{9}r^3; \ \ f(r) = 1; \ \ \beta(r) = \beta_0. \label{functionsEds}
\end{equation}
The new expressions for the scale factor $ R(r,t) $ and its derivatives are:
\begin{equation}
	R(r,t) = r(t + \beta_0)^{2/3}, \ \ \dot{R}(r,t) = \frac{2}{3}r(t + \beta_0)^{-1/3}, \label{Reds}
\end{equation}
\begin{equation}
	R'(r,t) = (t + \beta_0)^{2/3}, \ \ \dot{R}'(r,t) = \frac{2}{3}(t + \beta_0)^{-1/3}. \label{R'eds}
\end{equation}
The above equations define the EdS cosmology. The line element for this case is
\begin{equation}
 \mathrm{d}S^2 = \mathrm{d}t^2 - (t + \beta_0)^{4/3}[\mathrm{d}r^2 - (r)^2(\mathrm{d}\theta^2 + \sin^2{\theta} \mathrm{d}\phi^2)],
\end{equation}
Bearing in mind that the aim here is to analyze the model along the observer's past null cone, we can make use of the null geodesic equation,
\begin{equation}
	\frac{\mathrm{d}t}{\mathrm{d}r} = - \frac{R'}{f},
\end{equation}
and substitute the equations (\ref{functionsEds}), (\ref{Reds}), and (\ref{R'eds}) in the above expression and integrate it from $ t = 0 $, $ r = 0 $ (our ``here and now", or the observer's position) to $ t[r(\lambda)]$ to find,
\begin{equation}
	3(t + \beta_0)^{1/3} = 3\beta_0^{1/3} - r. \label{geodesicEdS}
\end{equation}
This relates the radial and temporal coordinates $ r $ and $ t $, and this equation can be used to rewrite the expressions (\ref{Reds}) and (\ref{R'eds}) along the past null cone as,
\begin{equation}
	R = \frac{r}{9}(3\beta_0^{1/3} - r)^2, \ \ \dot{R} = 2r(3\beta_0^{1/3})^{-1}, \label{Rdoteds}
\end{equation}
\begin{equation}
	R' = \frac{1}{9}(3\beta_0^{1/3} - r)^2, \ \ \dot{R'} = 2(3\beta_0^{1/3})^{-1}.
\end{equation}
Substituting the above expressions into equation (\ref{der_I}), and integrating it from $ I = 0 $, $ r = 0 $ to $ I(r) $, we find,
\begin{equation}
	1 - I = \left(\frac{3\beta_0^{1/3} - r}{3\beta_0^{1/3}}\right)^2,
\end{equation}
and now substituting it back into equation (\ref{redshiftI}), we obtain an expression for the redshift,
\begin{equation}
	1 + z(r) = \left(\frac{3\beta_0^{1/3}}{3\beta_0^{1/3} - r}\right)^2. \label{zreds}
\end{equation}
The equation (\ref{geodesicEdS}) can be used to obtain a value for the radial coordinate corresponding to a specific value of the temporal coordinate $t_B$, which is,
\begin{equation}
	r_B = 3[(t_B + \beta_0)^{1/3} - \beta_0^{1/3}]. \label{geodesicEdSrb}
\end{equation}
Finally, the radial coordinate and the luminosity distance are related by the following expression,
\begin{equation}
r = 3{t_B}^{1/3} - \left\{9^{1/2}{t_B}^{1/3}\left[\frac{1}{2} + \left(\frac{d_L}{3t_B} + \frac{1}{4}\right)^{1/2}\right]\right\}^{-1}. \label{rdlEDS}
\end{equation}




\chapter{The Single Radial Null Geodesic}\label{chapter3}

As seen in the last chapter, the inhomogeneous LTB model is a non trivial model when it comes to finding analytical solutions for the observational quantities that are generally of interest in the cosmological studies. In this sense, the idea of using a single null geodesic is a way of simplifying the problem of finding analytical solutions for observational quantities. While treating observations of the sky, we mostly deal with a single event on a cosmological scale, which means studying the observer's single past null cone. The concept of a single radial null geodesic translates that observational aspect into a geometrical one. For simplicity in the calculations and applications, we shall use the parabolic LTB solution as given by equation (\ref{paraB}), and a simultaneous big bang ($\beta_0 = constant$). In section 2.1 we present the concept of the single past null geodesic and in section 2.2 we derive the analytical expression for the redshift.

\section{The Concept}\label{concept}

\begin{figure}[!h]
\centering
\includegraphics[scale=0.75]{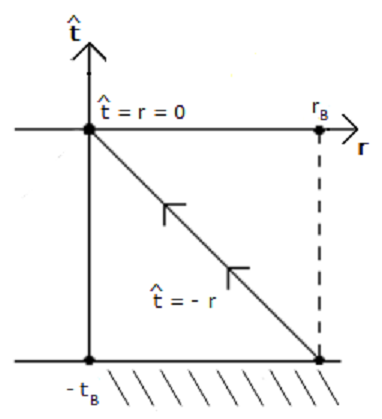}
\caption[Spacetime diagram for the past null geodesic]{Spacetime diagram for the past null geodesic represented by equation (\ref{uniquegeodesicint}), with $ \tau = 0 $. The Big Bang surface is given by  $ \hat{t} = - t_B $ which corresponds to a value $ r_B$ for the radial coordinate.}
\label{geodesic2}
\end{figure}

Mustapha et al. (1997) proposed the concept of a single null geodesic in order to find a particular solution for an observer's past null cone by assuming $ \mathrm{d}S^2 = 0 = \mathrm{d}\theta^2 = \mathrm{d}\phi^2 $ in the geodesic equation (\ref{geodesicLT}), to obtain an equation of motion along the past null cone. The result is a relation between the coordinates $ t = \hat{t} $ and $ r $, given by
\begin{equation}
	\mathrm{d}\hat{t} = - \frac{R'[\hat{t}(r), r]}{f}\mathrm{d}r = - \frac{\hat{R}'}{f}\mathrm{d}r, \label{radialgeo}
\end{equation}
where $ R[\hat{t}(r), r] = \hat{R} $. Then they chose the particular single radial null geodesic given by,
\begin{equation}
	\frac{\hat{R}'}{f} = 1.
\end{equation}
With this choice, the equation (\ref{radialgeo}) becomes,
\begin{equation}
	\frac{d\hat{t}}{d\lambda} = - \frac{dr}{d\lambda}, \label{uniquecond}
\end{equation}
whose integration yields,
\begin{equation}
	\hat{t}(r) = \tau - r, \label{uniquegeodesicint}
\end{equation}
where $ \tau $ is a constant of integration. We can see a representation of this single null geodesic in the spacetime diagram of Figure \ref{geodesic2}.
Note that the singularity surface $t_B$ denotes the corresponding value of the radial coordinate $r_B$. Taking into consideration the single null geodesic hypothesis, the equations (\ref{fieldone}) and (\ref{fieldtwo}) can be written in the form
\begin{equation}
	\dot{\hat{R}}^2 - \frac{F}{2\hat{R}} = f^2 - 1, \label{newfieldone}
\end{equation}
\begin{equation}
	8\pi\hat{\rho}\hat{R}^2 = \frac{F'}{2\hat{R}}. \label{newfieldtwo}
\end{equation}
We will now consider the hypothesis of a simultaneous big bang, or that $ \beta(r) = \beta_0 = t_B $, and choose the parabolic solution (\ref{paraB}) of the field equation (\ref{newfieldone}). The parabolic solution and its derivatives, with $\beta ' = 0$, are written as,
\begin{equation}
	\hat{R}(r,t) = \frac{1}{2}(9F)^\frac{1}{3}(\hat{t} + t_B)^\frac{2}{3},\label{Rponto}
\end{equation}
\begin{equation}
	\dot{\hat{R}} = \left[\frac{F}{3(\hat{t} + t_B)}\right]^{1/3}, \label{Rpt}
\end{equation}
\begin{equation}
	\hat{R'}(r,t) = \frac{1}{3}\left[\frac{9F}{(\hat{t} + t_B)}\right]^\frac{1}{3}\left[\frac{(\hat{t} + t_B)F'}{2F}\right], \label{Rlinha}
\end{equation}
\begin{equation}
	\dot{\hat{R'}}(r,t) = \frac{1}{9}\left(\frac{9F}{\hat{t} + t_B}\right)^\frac{1}{3}\left(\frac{F'}{F}\right). \label{Rpontolinha}
\end{equation}

\section{Analytical Redshift Expression}\label{analytic}

We can now obtain an analytical expression for the redshift in the LTB geometry for the parabolic solution using the single null geodesic equation (\ref{radialgeo}). In chapter \ref{chapterone}, section 1.2, we found an expression that relates the redshift $z$ and the radial coordinate $r$, given by equation (\ref{der_I}). Substituting the derivatives for the parabolic solution found in the previous section in equation (\ref{paraB}) and considering equations (\ref{uniquecond}) and (\ref{uniquegeodesicint}) we find the following differential equation,
\begin{equation}
\mathrm{d}I= \frac{2}{3}\frac{1}{(\hat{t} + t_B)} d\lambda. \label{intI}
\end{equation}
If we substitute the integration of the expression above into equation (\ref{finallag1}), we have as result the following expression, 
\begin{equation}
	\frac{\mathrm{d}\hat{t}}{\mathrm{d}\lambda} = \frac{1}{\displaystyle\int \frac{2}{3}\frac{1}{(\hat{t} + t_B)} \mathrm{d}\lambda - 1}. \label{eqdift1}
\end{equation}
This expression cannot be directely integrated because we do not know the behavior of the function $ \hat{t}(\lambda)$. So, we will approach the problem from another point of view. Considering that $ \hat{t} = 0 $ at $ r = 0 $ (our ``here and now") in the integrated single null geodesic equation (\ref{uniquegeodesicint}) yields that $ \tau = 0 $. Then, the equation (\ref{uniquegeodesicint}) turns to out to be given as,
\begin{equation}
	\hat{t} = - r. \label{tandr}
\end{equation}

\begin{figure}[!b]
\centering
\includegraphics[scale=0.9]{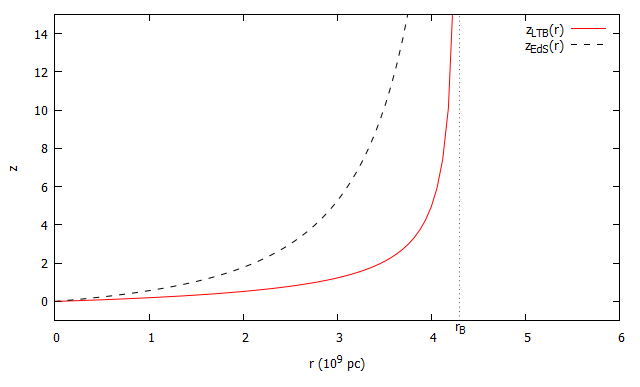}
\caption[The redshift as a function of the radial coordinate for the parabolic LTB and the Einstein-de Sitter models]{The redshift as a function of the radial coordinate for the parabolic LTB model (full line), assuming the single null geodesic hypothesis, and for the Einstein-de Sitter model. Considering $ c = G = 1 $, the temporal unit is expressed in $3.26\times 10^9$ years. The assumed value for the bang singularity is $ t_B $ = 4.3, which is based on the minimum estimated value of the age of the Universe by the $\Lambda$CDM cosmology, $\sim 14\times 10^9$ years. The redshift tends to infinity at $ r_B = 4.3\times 10^9 $pc (vertical dotted line). For the LTB model, the behavior $r > r_B $ has no physical meaning, because it defines the limit of validity of the model.}
\label{ZR}
\end{figure}

We wish now to rewrite the equation (\ref{int_I}) as a function of the radial coordinate $r$, instead of the affine parameter $\lambda$. Using the chain rule of derivatives, equation (\ref{int_I}) becomes,
\begin{equation}
	\frac{dI}{dr} = \frac{d}{dr} \int \frac{\dot{R}'}{R'}d\lambda = \frac{d}{d\lambda}\frac{d\lambda}{dr}\int \frac{\dot{R}'}{R'}d\lambda = \frac{d\lambda}{dr}\frac{\dot{R}'}{R'}.
\end{equation}
Substituting equation (\ref{der_I}) in the expression above and considering the single null geodesic hypothesis, we find
\begin{equation}
	\frac{dI}{dr} = \frac{(1 - I)}{\hat{R'}}\dot{\hat{R'}}.
\end{equation}
Now substituting the expressions (\ref{Rlinha}) and (\ref{Rpontolinha}) for the derivatives of $\hat{R}(t,r)$, we have as result the following differential equation,
\begin{equation}
\frac{dI}{dr} = \frac{2}{3}\left(\frac{1 - I}{\hat{t} + t_B}\right),
\end{equation}
which can be rewritten using the equation (\ref{tandr}), to find
\begin{equation}
	\frac{dI}{(1 - I)} = \frac{2}{3}\left(\frac{dr}{t_B - r}\right).
\end{equation}
Integrating the expression above once, we obtain as a result
\begin{equation}
	I(r) = 1 - C_3(t_B - r)^{2/3}.
\end{equation}
where $C_3$ is an integration constant. This expression can be transformed in an equation relating the redshift and the radial coordinate $r$, by simply using the relation given by equation (\ref{redshiftI}), which results in
\begin{equation}
	z(r) = \frac{1 - C_3(t_B - r)^{2/3}}{C_3(t_B - r)^{2/3}} = \frac{1}{C_3(t_B - r)^{2/3}} - 1. \label{redshiftfinal}
\end{equation}
If we apply the condition where $ z = 0 $ at $ r = 0 $, we obtain the constant $C_3$ to be equal to,
\begin{equation}
	C_3 = t_B^{-2/3}.
\end{equation}
The final expression for the redshift as a function of the radial coordinate is then written as,
\begin{equation}
	1 + z(r) = \frac{t_B^{2/3}}{(t_B - r)^{2/3}}. \label{redshiftfinal2}
\end{equation}
We can see the behavior of the function $z(r)$ in both the LTB and EdS cosmologies in Figure \ref{ZR}.

Figure \ref{ZR} shows that, for a given value of the comoving radial coordinate, we can obtain different values for the redshift, depending on the geometry considered. This is an interesting result, as measures of redshifts, and consequently distances, are often used to obtain constraints in cosmological models, as it is done with the SNIa measures and the $\Lambda$CDM cosmology.

With the expression (\ref{redshiftfinal2}) we can now obtain analytical expressions for other useful observational quantities, such as the number counts of sources and the global volume density. These will be obtained in the next chapter, in the context of a fractal approach for modelling the matter distribution of the Universe.




\chapter{Fractals and the LTB cosmology}\label{chapter4}

In this chapter we apply the single null geodesic approach for the LTB spacetime to model the observable universe as a fractal distribution. The single null geodesic technique allows us to go further than Ribeiro (1992a, 1993) in modeling a fractal distribution with the LTB geometry, as we can now carry out an analytical treatment rather than a numerical one. Some authors have recently  been modelling the observable universe by dividing it into regions having different fractal dimensions (Sylos Labini et al. 2008, Sylos Labini 2011b), and as we shall see further in this chapter, this viewpoint is also present in the results obtained by our theoretical fractal approach. The first section shows how the concepts of hierarchical structures and self-similarity have been slowly developed by many authors throughout history and became finally synthesized in the concept of fractals. In the section 3.2 we use the analytical solutions obtained in section 2.2 to derive a simple fractal LTB model and in the last section we present and discuss the results.
\newpage
\section{Fractals}
\begin{quotation}
\begin{flushright}
``Theories crumble, but good observations never fade.''\\(H.\ Shapley)
\end{flushright}
\end{quotation}
\subsection{Historical Background}

The use of fractals as a practical tool to model complex systems is mainly credited to Mandelbrot, who proposed in his book \textit{The Fractal Geometry of Nature} (1982) that fractals can describe several very irregular shapes found in nature. That approach was followed by many authors in different fields of science, such as biology, fluid mechanics, geo-statistics, economics, telecommunications and other areas. But the concept of fractals, including its use in the cosmological context, was already being developed much earlier by many authors under other terminologies and different aspects. We will briefly describe here some of these works. For a more detailed discussion, see Ribeiro (1994, 2005), Conde-Saavedra (2011) and Conde-Saavedra et al. (2013).

In this sense, the earlier concept of hierarchical structures can already be found in the works of E. Swedenborg (1734), I. Kant (1724 - 1804) and J. H. Lambert (1728 - 1777). Even in the presocratic Greece, the philosopher Anaxagoras (500 - 428 B. C.) presented a theory of how matter seemed to be divisible without limit and grouped under the same structures in various scales, an idea similar to the self-similarity concept later derived by other authors. Kant thought that the idea of hierarchical systems could be used to explain the stellar and galactic systems, and even wrote that there would be an infinite progression of worlds and systems, or that there is an infinite hierarchy for these systems, differently of Swedenborg, who proposed in his book \textit{Principia} that the systems were hierarchically organized but not \textit{ad infinitum} (Baryshev \& Teerikorpi 2005).

John Herschel (1792 - 1871) also thought there could be an infinite universe when he was searching for an explanation of the Olber's Paradox, which would allow many structures to form and that there was an infinite amout of directions so that we could look at one and not find a star. For that, he concluded that there were many systems that obey a law that ``every higher order of bodies should be immensely more distant from the centre than those of the next inferior order", which could lead to some sort of hierarchical system, or, as he called, a ``cosmic law" (c.f. Baryshev \& Teerikorpi 2005).

The idea of hierarchical structure of the universe was finally put into a practical and mathematical frameworks in 1907 by Edmund Fournier d'Albe (1868 - 1933). In his work \textit{Two new worlds}, stars were distributed in a hierarchy of spherical clusters in an infinite space, and the mass inside each sphere would increase proportionally to its radius,
\begin{equation}
M(R) \sim R.
\end{equation}
This idea of his was born as an explanation for the cosmological paradoxes in Newton's universe.

The astronomer Carl Charlier (1862 - 1934) published an article in 1908 called \textit{How an infinite world may be built up} where he studied Fournier d'Albe's ideas to develop more general models of stellar distributions. He found a criterion which hierarchy would have to fulfill so that it could solve the Olber's Paradox and the infinite gravitational potential as well. He proposed that the important factor was how fast the density would increase from one level (i) to the next one (i + 1), and that this would depend on the ratio of the sizes of sucessive elements and on the number $N_{i+1}$ of the lower elements forming the upper one. Its criterion can be written as
\begin{equation}
\displaystyle \frac{R_{i+1}}{R_i} \geq N_{i+1},
\end{equation}
where $R_i$ and $R_{i+1}$ are the sizes (radia). He later derived a second criterion,
\begin{equation}
\displaystyle \frac{R_{i+1}}{R_i} \geq \sqrt{N_{i+1}},
\end{equation}
after a note from Selety in 1922. In terms of continuous mass-radius behaviour for identical particles of mass $m$, the first criterion corresponds to $M(R) = mN(R) \sim R$, and the second gives $M(R) \sim R^2$, which is sufficient to solve the Olber's paradox and the infinite gravity force.

The concept of hierarchical clustering was also investigated by Carpenter (1938), from data obtained at the Harvard and Mount Wilson observatories. He found that the smaller clusters were denser than the bigger ones, and, therefore, galaxies were distributed in a nonuniform, though not random way. As his work lacked a mathematical formulation, his results and the hierarchical cosmology were abandoned for some time.

In 1970 G. de Vaucouleurs proposed a hierarchical cosmology model where the galaxies would follow an average global density power law with negative slope, of the form $\bar{\rho} \propto r^{-1.7}$. Soon after, J. R. Wertz (1971) studied some possible models of a Newtonian hierarchical cosmology, and then proposed some observational tests for these models (see Ribeiro 1994 and references therein).

These specific works were synthesized in the idea of fractals, proposed by Mandelbrot in 1977. He introduced the term and gave its definition: ``\textit{fractal is a set for which the Hausdorff dimension strictly exceeds the topological dimension}", though asserting that such a definition for fractals would problably be too restrictive. Later, he relaxed this definition and stated that a fractal can be described as ``a shape made of parts similar to the whole in some way", which is essentially the concept of hierarchical clustering where smaller systems like galaxies can group together to form bigger systems, like clusters and superclusters, which would be structures with similar characteristics but at larger scales (see Ribeiro 1994). 

To clarify the concept of a fractal dimension associated to irregular structures, let us take as an example the classic coastline length problem (Mandelbrot 1967). A coastline is made of self-similar parts, so that more and more structures are covered if we reduce to smaller units, which lead to an increase in the length of the coastline. Then, the smaller the unit considered, the higher the precision, and longer the coastline will be, reaching a limit of an infinite perimeter. So the precision of the length measure is dependent on the covering units. To solve this contradiction, Mandelbrot proposed that the coastline perimeter should be expressed by a power law, 
\begin{equation}
	L(\delta) \propto \delta^N(\delta),
\end{equation}
where $N(\delta) \propto \delta^{-D}$ is the number of $\delta$ scales that covers the coastline, and $D$ is the fractal dimension, which denotes the level of regularity the coastline. The higher the value of the fractal dimension, the more irregular the coastline is. The measured value for the fractal dimension of the Great Britain coastline is $D = 1.31$, less irregular than the coast of Norway, found to have $D = 1.452$. The non-integer values denote that the coastlines are somewhere between straight lines ($D = 1$) and surfaces ($D = 2$).

Other authors, such as W. B. Bonnor (1972) and P. S. Wesson (1978-1979) studied relativistic models with density power laws, with similar exponents ($\bar{\rho} \propto r^{-1.7}$ and $\bar{\rho} \propto r^{-2}$, respectively). In 1987 L. Pietronero published a model where the large-scale distribution of galaxies formed a single fractal system, obtaining a de Vaucoulers' type global density power law $\bar{\rho} \propto r^{-\gamma}$ as result. In 1988, R. Ruffini, D. J. Song \& S. Taraglio proposed that the fractal system should have an upper cutoff to homogeneity in order to solve what they called as an apparent conflict between ``the commonly accepted idea in theoretical cosmology
that greater distances represent earlier epochs of the Universe implying that higher
average densities should be observed", a transition that was suggested by Wertz (1970). This contradiction was found to be false (Ribeiro 2001, 2005; Rangel Lemos \& Ribeiro 2008), as the use of different distance definitions implies in different behaviors of the local and average densities. As $z \rightarrow \infty$, the local density tends to infinity, but the average density along the past null cone tends to zero. Other hierarchical systems were also studied by E. Abdalla $et.$\ $al$ (1998, 1999, 2004), who also found density distributions along the past null cone which are described by power laws in spatially homogeneous models.

\subsection{Mathematical Development of a Simple Fractal Model}

Now we will develop the hypothesis of self-similarity in a more practical way. We can say that the idea of self-similarity implies that the rescaling of the length $r$ by a factor $b$ (Baryshev \& Teerikorpi 2005),
\begin{equation}
r \rightarrow \tilde{r} = br,
\end{equation}
leaves that some property, presented by an arbitrary function $s(r)$ remains unchanged, apart from a renormalization that
depends on $b$, but not on the variable $r$. This leads to the functional relation
\begin{equation}
s(\tilde{r}) = s(b\cdot r) = A(b) \cdot s(r),
\end{equation}
which is satisfied by a power law with any exponent. For
\begin{equation}
s(r) = s_0r^d,
\end{equation}
we have
\begin{equation}
s(\tilde{r}) = s_0(br)^d = (b)^d s(r).
\end{equation}
The exponent $d$ defines the behaviour of the function everywhere, and if we construct a self-similar model or a fractal structure using this function, this exponent will have a fundamental physical meaning of a fractal dimension. If we assume the condition for the amplitude $s(r_0) = 1$, this implies a certain length $s_0$ given by,
\begin{equation}
r_0 = s_0^{-1/d}.
\end{equation}
Note that this is not a characteristic length because the power law was constructed as self-similar at the beginning. Thus, there is no preferred scale as the exponent $d$ is a constant dimensionless factor and is not related to any length scales.

The single fractal model presented by L. Pietronero (1987) applies this mathematical point of view to a deterministic fractal structure for the galaxy distribution. 
\begin{figure}[!ht]
\centering
\includegraphics[scale=0.85]{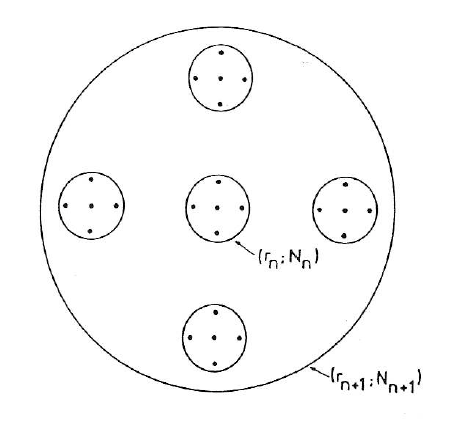}
\caption[A schematic illustration of a deterministic
fractal system from where a fractal dimension can be derived]{A schematic illustration of a deterministic
fractal system from where a fractal dimension can be derived. The structure
is self-similar, repeating itself at different scales (Pietronero 1987).}
\label{hierfrac}
\end{figure}
The basic idea of his model can be derived as follows. Starting from a point occupied by an object, we count how many objects are present within a volume characterized by a certain length scale, to get $N_0$ objects within a
radius $ r_0$, $N_1 = \tilde{k}N_0$ objects within a radius $r_1 = kr_0$ and $N_2 = \tilde{k}N_1 = \tilde{k}^2 N_0$ objects within a radius $r_2 = kr_1 = k^2r_0$, as illustrated by Figure \ref{hierfrac}. In general we have
\begin{equation}
N_n = \tilde{k}^nN_0, \label{Nfractal}
\end{equation}
within
\begin{equation}
r_n = k^nN_0, \label{rfractal}
\end{equation}
where $\tilde{k}$ and $k$ are constants. If we take the logarithms of the equations (\ref{Nfractal}) and (\ref{rfractal}) and divide one by the other we get,
\begin{equation}
N_n = \sigma(r_n)^D, \label{nfractaldiscrete}
\end{equation}
with 
\begin{equation}
\sigma \equiv \frac{N_0}{(r_0)^D}, \label{sigma0}
\end{equation}
\begin{equation}
D \equiv \frac{\log \tilde{k}}{\log k},
\end{equation}
where $\sigma$ is a prefactor of proportionality related to the lower cutoffs $N_0$ and $r_0$ of the fractal system, i.e., the inner limit of the fractal system, and $D$ is the fractal dimension. 

With this simple model in mind, in the next section we will derive a fractal model in the LTB space-time.

\section{A Fractal Approach for the LTB Cosmology}\label{approach}

\indent We will now derive the expressions that relate the LTB geometry to a simple fractal model. Following Pietronero's assumption about the fractal distribution of cosmic objects, the expression for the number of these objects $ N $ inside a spherical region whose radius is given by the luminosity distance $ d_{L} $ as follows,
\begin{equation}
	{N} = \sigma(d_{L})^D. \label{nfractal}
\end{equation}
The global density of a fractal distribution having dimension $D$ is given by the de Vaucoulers' power law, as below,
\begin{equation}
	\rho_L = \frac{N \mathcal{M}_g}{V_L} = \frac{3\sigma \mathcal{M}_g}{4\pi}(d_L)^{(D - 3)}. \label{densityfrac}
\end{equation}
The two expressions for the global density, equations (\ref{densityLT}) and (\ref{densityfrac}) are the same quantity, so they can be written as the following expression,
\begin{equation}
	\int_C \frac{F'}{f} dr = 4\sigma \mathcal{M}_g[\hat{R}(1 + z)^2]^D.
\end{equation}
This expression can be called as the self-similarity condition for the LTB model in the single null geodesic regime. This condition parametrizes the LTB matter distribution and curvature in terms of the two parameters which characterize a single fractal distribution ($\sigma, D$).

Substituting in the above the expression the equation (\ref{redshiftfinal2}) for the redshift and equation (\ref{Rponto}) for $\hat{R}$, we have,
\begin{equation}
	\int_C \frac{F'}{f} dr = 4\sigma \mathcal{M}_g\left[\frac{1}{2}(9F)^{1/3}(t_B - r)^{2/3}\left(\frac{t_B^{2/3}}{(t_B - r)^{2/3}}\right)^2\right]^D.
\end{equation}
If we assume now $f(r) = 1$, i.e., the model has the same positive constant curvature for any value of $r$, we obtain an expression for the function $F(r)$, which obeys a fractal matter distribution as follows,
\begin{equation}
F = 4\sigma \mathcal{M}_g\left[\frac{1}{2}(9F)^{1/3}(t_B - r)^{2/3}\left(\frac{t_B^{2/3}}{(t_B - r)^{2/3}}\right)^2\right]^D. \label{FrC}
\end{equation}
The above equation can be rewritten as,
\begin{equation}
F(r) = \left(4\sigma
\mathcal{M}_g\right)^{3/(3-D)}\left\{\frac{1}{2}(9)^{1/3}\left[\frac{t_B^{4/3}}{(t_B - r)^{1/3}}\right]\right\}^{3D/(3-D)}. \label{FR}
\end{equation}
Using the expression (\ref{redshiftfinal2}) we can rewrite equation (\ref{FR}) as a function of the redshift $z$ as below,
\begin{equation}
F(z) = \left(4\sigma
\mathcal{M}_g\right)^{3/(3-D)}\left[\frac{1}{2}(9)^{1/3}\left(\frac{t_B^{2/3}}{1
+ z}\right)(1 + z)^2\right]^{3D/(3-D)}. \label{FZ}
\end{equation}

We can compare the behavior of the global density for Einstein-de Sitter and parabolic LTB models. Substituting the equations (\ref{Rponto}) and (\ref{redshiftfinal2}) in the equation (\ref{densityLT}), along with the assumption that $f(r) = 1$, we obtain the global density in the parabolic LTB model, given by,
\begin{equation}
	{\gamma^{*}}_{LTB}(r) = \frac{(t_B - r)^{2}}{6\pi {t_B}^4}. \label{rhoLTB}
\end{equation}
For the Einstein-de Sitter case, we use equations (\ref{Rdoteds}), (\ref{functionsEds}) and (\ref{zreds}) in equation (\ref{densityLT}) to obtain the following expression,
\begin{equation}
	{\gamma^{*}}_{EdS}(r) = \frac{(3t_B^{1/3} - r)^6}{54\pi(3t_B)^4}. \label{rhoeds}
\end{equation}
Substituting the expression (\ref{rdlEDS}) in equation (\ref{rhoeds}), we come up with the expression for the global density as a function of the luminosity distance, that is,
\begin{equation}
	{\gamma^{*}}_{EdS}(d_L) = \frac{1}{6\pi{t_B}^2}\left(\frac{1}{2} + \sqrt{\frac{d_L}{3t_B} + \frac{1}{4}}\right)^{-6}. \label{rhoedsdL}
\end{equation}
Note that these expressions for the global densities do not depend on the parameters $\sigma$ and $D$, because the function $F(r)$ cancels out during their derivation.
\section{Results}\label{results}

\indent We ran a few tests of the model for various intervals of redshift and the radial coordinate. The results can be seen in the graphs below. We compare the LTB model with the Einstein-de Sitter one. 

\begin{figure}[!t]
\centering
\includegraphics[scale=0.85]{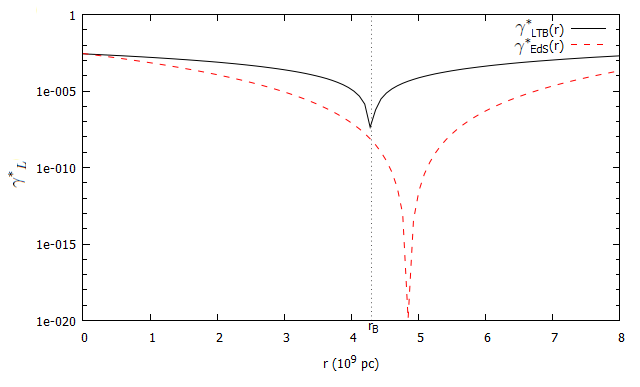}
\caption{Luminosity distance global density $ vs. $ the radial coordinate for the parabolic LTB and the Einstein-de Sitter models.}
\label{plotRHOR}
\end{figure}

The objective is to obtain the parameters $\sigma$ and $D$ which generate a fractal distribution of matter such that $F = F(\sigma, D)$ and compare it with the distribution of matter in the Einstein-de Sitter cosmology. Fig.\ \ref{plotRHOR} shows the global densities for the parabolic LTB and the Einstein-de Sitter cosmologies in terms of the radial coordinate $r$ (not dependent on $\sigma$ and $D$), using the expressions (\ref{rhoLTB}) and (\ref{rhoeds}), respectively. We see that the global density for the LTB cosmology reaches a minimum value faster than for the Einstein-de Sitter cosmology. This happens because the singularity of the parabolic LTB global density is defined directly by the value assumed for $t_B$, and in the Einstein-de Sitter case the singularity of its global density is $3t_B^{1/3}$, as stated by equation (\ref{rhoeds}).

\begin{sidewaysfigure}[h]
\centering
\includegraphics[scale=1.3]{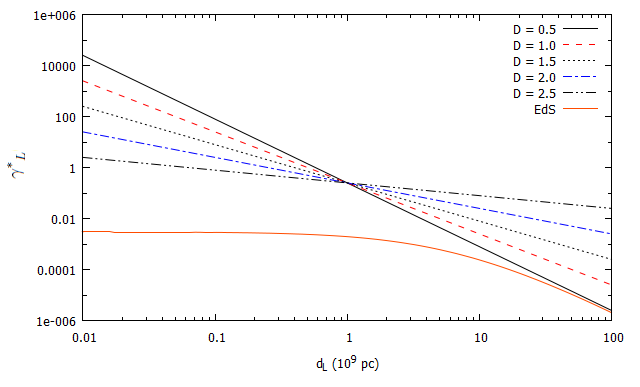} 
\caption[Global density $ versus $ the luminosity distance, for different values of the fractal dimension $ D $ and for the Einstein-de Sitter model]{Global density $ versus $ the luminosity distance, for different values of the fractal dimension $ D $ and for the Einstein-de Sitter model. The assumed value for $ \sigma $ is the unity for all fractal dimensions. For high values of the luminosity distance, the model with $ D = 0.5 $ approximates the Einstein-de Sitter distribution. So, for high values of $d_L$, a parabolic LTB model with $\sigma = 1, D = 0.5$ approximates the Einstein-de Sitter cosmology.}
\label{plotRHODLEDS}
\end{sidewaysfigure}

\begin{sidewaysfigure}[h]
\centering
\includegraphics[scale=1.3]{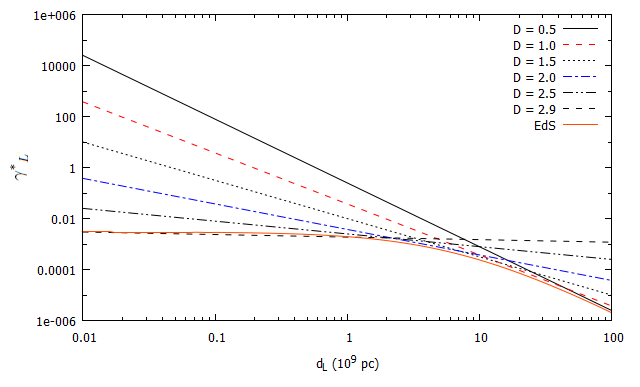} 
\caption[Global density $ versus $ the luminosity distance, for different values of the fractal dimension $ D $ and for the Einstein-de Sitter model]{Global density $versus$ the luminosity distance, for different values of the fractal dimension $ D $ and for the Einstein-de Sitter model. The assumed values for $ \sigma $ now differ for every fractal dimension ($ D = 0.5 $, $\sigma = 1 $; $ D = 1 $, $\sigma = 0.15 $; $ D = 1.5 $, $\sigma = 0.04 $; $ D = 2 $, $\sigma = 0.015 $; $ D = 2.5 $, $\sigma = 0.01 $; e  $ D = 2.9 $, $\sigma = 0.0075 $). For different values of the luminosity distance, a certain fractal model, which corresponds to a ($D, \sigma$) parametrized LTB model, can be approximated as the Einstein-de Sitter cosmology. From the high values to the smaller ones, the fractal dimension that corresponds to the Einstein-de Sitter model gets higher, until it would eventually reach $ D = 3 $.}
\label{plotRHODLEDSSIGMA}
\end{sidewaysfigure}

\begin{sidewaysfigure}[h]
\centering
\includegraphics[scale=1.3]{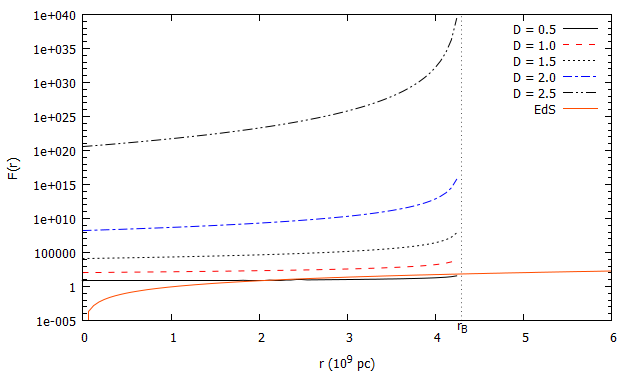} 
\caption[Graph of the function $ F(r) $ for different values of the fractal dimension $ D $, compared to the Einstein-de Sitter model]{Graph of the function $ F(r) $ for different values of the fractal dimension $ D $, compared to the Einstein-de Sitter model. The assumed value for $ \sigma $ is the unity for all fractal dimensions. The vertical dotted line denotes the singularity of the model at $ r_B = 4.3\times 10^9 $pc. Note that the fractal distribution with $ D = 0.5 $ is approximately the Einstein-de Sitter distribution in a certain range of the radial coordinate $r$.}
\label{plotfrALLDEdS}
\end{sidewaysfigure}

\begin{sidewaysfigure}
\centering
\includegraphics[scale=1.3]{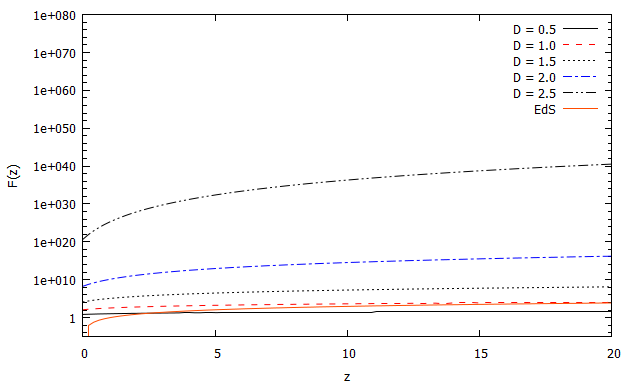} 
\caption[Graph of the function $ F(z) $ for different values of the fractal dimension $ D $, compared to the Einstein-de Sitter model]{Graph of the function $ F(z) $ for different values of the fractal dimension $ D $ and the Einstein-de Sitter model. The assumed value for $ \sigma $ is the unity for all fractal dimensions.} 
\label{plotFZALLDEDS}
\end{sidewaysfigure}

In Fig.\ \ref{plotRHODLEDS} we plot the global density as a function of the luminosity distance for the fractal and the Einstein-de Sitter models, which are given by the expressions (\ref{densityfrac}) and (\ref{rhoedsdL}), respectively, and considering the lower cutoff constant $\sigma$ as the unity for any value of the fractal dimension. We see that the fractal models shows a linear behavior for any fractal dimension, and that the Einstein-de Sitter is linear (constant) only in a certain range of values of the luminosity density, eventually decreasing for higher values. In Fig.\ \ref{plotRHODLEDSSIGMA} we changed the values of the constant $\sigma$, for different fractal dimensions. From this plot we can see that, for each different fractal dimension, the value of $\sigma$ can be adjusted to match the Einstein-de Sitter model in different ranges of the luminosity distance. 
Figures \ref{plotfrALLDEdS} and \ref{plotFZALLDEDS} show the behavior of the functions defined by equations (\ref{FR}) and (\ref{FZ}) which can be interpreted as the cumulative distribution of matter inside a radius $r$ or a redshift $z$. We see that the distribution grows with $r$ and eventually reaches a maximum value at the singularity of the model. For the redshift plot, it grows indefinitely.

In all graphs presented here we assume $\mathcal{M}_g = 1$, which means that we always consider the cosmological sources to have the same mass, equal to $2.09\times 10^{22} M_{\odot}$.



The plots indicate that the parabolic LTB fractal model generally tends to the Einstein-de Sitter one for high values of the luminosity distance, radial coordinate and the redshift, for low values the fractal dimension. This indicates that, as $z \rightarrow \infty$, the fractal model assymptotically tends to the Einstein-de Sitter cosmology if the fractal dimension is low.






\chapter*[Conclusions]{Conclusions}
\addcontentsline{toc}{chapter}{Conclusions}
\label{chapter5}

\indent In this work we have presented an overview of the spatially inhomogeneous Lema\^itre-Tolman-Bondi geometry, including its derivation from the field equations of General Relativity. We used the concept of the single past null geodesic to obtain analytical solutions for many relevant observational quantities, such as the redshift and the number counts of cosmic objects. These analytical solutions allowed us to study a simple fractal model in the context of the parabolic LTB inhomogeneous geometry, avoiding then a numerical approach, as was done by Ribeiro (1993).

In chapter \ref{chapterone} we presented the inhomogeneous solution and discussed some of the properties of the model, showing two commonly used notations of the LTB geometry along with their interpretations. We also presented the observational quantities that are used in the context of cosmology and astrophysics, such as the area, luminosity and galaxy distances, the source number counts, the global density and finally the redshift, which we obtained by two different but equivalent methods. The chapter ended with a demonstration of the generality of the LTB geometry, as it allows the possibility of generating Friedmann models by a choice of the LTB arbitrary functions.

We presented the concept of the single radial null geodesic in chapter \ref{chapter3}, and used it to obtain an analytical solution of the parabolic LTB model for the redshift in terms of the radial coordinate. This solution allowed us to derive expressions for observational quantities of interest in this particular cosmology, such as the number counts and the global density in an entirely analytical manner as opposed to the numerical approach of Ribeiro (1992a, 1993). This work was motivated by a lack of analytical fractal solutions for the LTB model until this date.

With the expressions obtained in chapter \ref{chapter3} we derived a simple fractal model in chapter \ref{chapter4}, along with the results and interpretations of the functions of the fractal model. A comparison with the Einstein-de Sitter geometry was made to clarify the interpretations and discussions of this fractal model. The conclusions we obtained are as follows:

\begin{itemize}
\item For high values of the luminosity distance ($\sim 10^2$ to $10^3$ Gpc), the fractal global density tends to the Einstein-de Sitter one if the fractal dimension nears $D = 0.5$, if we use the same value of the lower cutoff constant $\sigma = 1$ for all the fractal dimensions. This behavior means that low values of the fractal dimension $D$ imply an equivalent behavior of the global density in both the spatially homogeneous model and a simple fractal model with $D = 0.5$ and for high values of the luminosity distance.
\item For different values of the lower cutoff constant $\sigma$ we find that the global density for the fractal model can be approximated to the Einstein-de Sitter's, depending on the assumed values of the luminosity distance and the fractal dimension. This behavior indicates that a change in $\sigma$, which is related to the initial values of the number of objects and the radius of the fractal system, can significantly alter the behavior of the global density of a fractal system.
\item The behavior of the function $F(r)$, which denotes the cumulative distribution of matter in a given radius, and is described by equation (\ref{FR}), indicates that, for $D \sim 0.5$, the distribution of the parabolic LTB model tends to the Einstein-de Sitter's distribution, for $r > 2\times 10^9$pc. For any value of the fractal dimension $D$, the function reaches a maximum value near the singularity of the parabolic LTB model, which in this case is located at the radial coordinate $r_B = 4.3\times 10^9$pc.
\item The cumulative distribution of matter as a function of the redshift, described by equation (\ref{FZ}), also has a tendency to the Einstein-de Sitter distribution as the fractal dimension assumes low values ($0.5 < D < 1.0$) for high redshifts ($z \simeq 2$), a similar behavior seen the radial distribution in Fig. \ref{plotfrALLDEdS}.
\end{itemize}

This similar behavior of a tendency for the Einstein-de Sitter cosmology for high values of the luminosity distance, radial coordinate and the redshift, found in all of our results, indicate that the parabolic LTB and fractal models can reach a spatially homogenenous behavior in the larger scales. Nevertheless, we see that for every value of the fractal dimension the fractal global density of matter follows a linear pattern, related to a power-law, which is independent of the scale considered. All the fractal distributions ``intersect" in a certain small range of values of the luminosity distance or the radial coordinate, when we consider the same value for the lower cutoff constant $\sigma$. This constant is interpreted as an initial distribution of objects in a certain radius. That being said, it may suggest an observational constraint of the model which could define the best suitable fractal dimension for the considered range of luminosity distance/redshift, and consequently constraining the parabolic LTB fractal model.
Following this viewpoint, the distribution of matter can be modeled with more than one value for the fractal dimension, depending on the scale considered, as we can have different values of the constant $\sigma$ for different scales. So, these indications of the variation of the fractal dimension with the luminosity distance/redshift mean that a spatially homogeneous distribution of matter (galaxies) can be approximated by an inhomogeneous LTB model at some scales with fractal features.

It is worth mentioning that this way of modelling the Universe, with the use of a inhomogeneous cosmology, does not necessarily mean that the whole Universe can be described as inhomogeneous. It may happen that this description holds only in a certain range of scales or redshift (Tolman 1934). 

The results of this work show that lower values of the fractal dimension approximate a spatially homogeneous distribution of matter along the past null cone, and that the lower cutoff constant $\sigma$ may be used as an observational and theoretical constraint to a fractal model. So, for higher values of the luminosity distance/redshift and lower densities, one should expect a lower fractal dimension model to be approximately a spatially homogeneous cosmology, which is the opposite result for smaller scales, in which a higher fractal dimension approximates a spatially homogeneous cosmology.

This work can be expanded as these parameters and observational quantities can be directly compared with observational results. A comparison with the standard model $\Lambda CDM$ can also be made in order to evaluate the observational concordance of these models.






\backmatter


\backmatter

\normalfont


\footnotesize
\printindex


\end{document}